\documentclass[12pt]{article}
\usepackage{geometry,xcolor}
\usepackage{fullpage} 
\usepackage{setspace}
\usepackage{enumitem}
\usepackage{appendix}
\usepackage{multirow}
\usepackage{threeparttable}
\usepackage{booktabs}
\usepackage[figuresright]{rotating}
\usepackage{natbib}
\usepackage{amsfonts,amsmath,mathrsfs,amssymb}

\usepackage{graphics}
\usepackage{graphicx}
\usepackage{subcaption} 
\usepackage[export]{adjustbox} 
\usepackage{ntheorem}
\usepackage[unicode]{hyperref}
\RequirePackage{algorithm}
\RequirePackage{algorithmic} 
\allowdisplaybreaks[4]
\hypersetup{
    colorlinks=true,
    linkcolor=blue,
    citecolor=blue, 
}

\newtheorem{theorem}{Theorem}[section]

\newtheorem{corollary}{Corollary}[section]

\newtheorem{assumption}{Assumption}
\newtheorem{case}{Case}

\graphicspath{ {./images/} }

\title{\textbf{Statistical Inference for Smoothed Support Vector Machines in High Dimensions: From Offline to Online Data}}
\begin{document}


\author{
  Shuya Zhou$^{1,2\dagger}$, Junwen Xia$^{1,2\dagger}$, and Jingxiao Zhang$^{1,2*}$ \\[2ex]
  {\small \it $^1$Center for Applied Statistics, Renmin University of China, Beijing, China} \\
  {\small \it $^2$School of Statistics, Renmin University of China, Beijing, China} \\[1.5ex]
  {\small \it $^\dagger$Equally contributing authors.} \\
  {\small \it $^*$email: zhjxiaoruc@163.com}
}

\date{} 
\maketitle



\begin{spacing}{1.2} 
\begin{abstract}
High-dimensional classification problems often rely on the Lasso-penalized linear Support Vector Machines (SVMs). However, the double non-smoothness induced by the hinge loss and Lasso penalty in this model makes statistical inference challenging and impedes computational efficiency. In this paper, we propose a unified inference framework in both offline and online settings. In the offline case, by applying a convolution smoothing technique to the hinge loss, we construct a debiased estimator that eliminates the shrinkage bias, thereby building a valid confidence interval. For online streaming data, we develop a real-time estimator and inference procedure that relies only on summary statistics of historical data. Theoretically, we provide rigorous proofs for the asymptotic normality of our offline and online debiased estimators. Simulation studies and real data applications demonstrate that our methods achieve valid statistical inference and improved computational efficiency.
\end{abstract}
\end{spacing}

\vspace{1em}

\newpage
\section{Introduction}

With the rapid advancement of data collection technologies, high-dimensional classification data are increasingly common, arising either as fixed offline datasets or as streaming online datasets in which observations arrive sequentially and statistical analysis must be performed in real time. Linear support vector machines (SVMs) are powerful tools for learning sparse and robust decision boundaries in high-dimensional feature spaces for both offline and online classification settings. Despite their popularity in practice, the resulting optimization problem is double nonsmooth: both the hinge loss and the Lasso penalty are nondifferentiable, which poses substantial computational challenges. As demonstrated in \cite{density-convoluted,wang2025convolution}, the computational cost of SVMs scales poorly as either the sample size or the feature dimension increases. 

To alleviate the computational burden, the smoothed SVM was introduced \cite[]{density-convoluted}. The smoothed SVM method smoothes the hinge loss function using the kernel function to obtain a differentiable smoothed hinge loss function and leverages the smoothed hinge loss to derive the linear decision boundaries. Due to the differentiability of the smoothed hinge loss, efficient algorithms tailored to a differentiable loss plus Lasso penalty can be leveraged \cite[]{parikh2014proximal}, thereby substantially relieving the concern of computation inefficiency. 
The smoothed SVM has now become a standard implementation to derive decision boundaries for high-dimensional classification data.

After the proposal of the smoothed SVM, substantial progress has been made in this field. \cite{wang2024convolution} incorporated nonconvex penalties, such as the SCAD penalty, into the smoothed SVM framework to achieve oracle properties in estimating the decision boundary. \cite{meng2025online} developed an online smoothed SVM to accommodate streaming data. \cite{wang2024svmsmoothing} proposed a distributed smoothed SVM, in which data are stored across local servers and a central computation server aggregates summary information from local servers to estimate the decision boundary. To further relax the reliance on a central server, \cite{chen2025efficient} introduced a decentralized smoothed SVM, where local servers communicate summary statistics only with their neighbors.

Despite the rapid advancement of the smoothed SVM, the statistical inference for the parameters of the decision boundaries in high dimensions for both offline and online datasets remains unsolved. The parameters of linear decision boundaries are crucial for model interpretation, thus, statistical inference for these parameters is necessary to identify which variables actually have real effects on this boundary. Motivated by this need, in this paper, we propose a smoothed debiased Lasso procedure for high-dimensional smoothed SVMs inference in an offline dataset, and an online updating smoothed debiased Lasso procedure for online dataset. We summarize our contributions as follows.

First, we propose a novel debiasing procedure for constructing confidence intervals of the decision boundary parameters in high-dimensional settings. Statistical inference in high dimensions is substantially more challenging than in low-dimensional cases because the Lasso penalty used for parameter estimation introduces bias. The problem is further complicated in smoothed SVMs, where the smoothing procedure induces additional bias. To simultaneously account for these two sources of bias, we develop a smoothed debiased Lasso procedure.

Second, we introduce a novel online updating debiasing method for conducting statistical inference in high-dimensional smoothed SVMs with streaming data. Compared to the offline setting, statistical inference for streaming data is more challenging due to storage constraints that prevent retaining all historical observations. To address this limitation, we develop a debiasing approach that combines summary statistics from previous data batches with individual observations from the current batch. Because the use of summary statistics introduces approximation error relative to the full dataset, our method further corrects for the additional bias induced by this approximation, beyond the bias addressed by offline debiased estimators.

Third, we provide rigorous theoretical guarantees for the proposed methods in both offline and online settings. Establishing asymptotic normality is particularly challenging in the smoothed SVM. Existing work on statistical inference for models such as smoothed quantile regression \cite[]{xie2025statistical} and smoothed rank regression \cite[]{WANG_rank_online} relies on the continuous residual, where the residual is defined as the difference between the response and its conditional mean. However, in SVM classification, the response is binary, so the corresponding residual is discrete, rendering the standard techniques in existing inferential works inapplicable. To overcome this challenge, we focus on the one-dimensional marginal contribution of covariates and introduce an empirical process technique that can transform the non-smooth empirical process into a Lipschitz-continuous one. Using this approach, we establish asymptotic normality for both the offline and online debiased estimators. En route to it, we first derive finite-sample error bounds for the Lasso estimators of smoothed SVMs under both offline and online settings, which may be of independent interest.

\subsection{Related literature}
In the offline setting, statistical inference for high-dimensional models has gained considerable attention in recent years. For high-dimensional linear models, \cite{zhang2014confidence,debias_lasso_vande, debias_lasso_javan} proposed the debiased Lasso procedure to remove the regularization bias incurred by the Lasso penalty, yielding asymptotically normal estimators.

Following the success in linear models, \cite{ning2017general} developed a unified theory to facilitate valid statistical inference for a broad class of high-dimensional M-estimators, including logistic and Poisson regression models.
Nevertheless, for models with non-smoothed loss functions, statistical inference remains challenging due to the inherent non-differentiability of the objective function. Statistical inference based on the non-smoothed loss typically deteriorates in both computational efficiency and the validity of inference \cite[]{Rybak_jmlr}. To address this issue, convolution smoothing has been employed to approximate the non-smooth loss function. For example, \cite{fernandes2021qrsmoothing, tan2022high} developed inferential frameworks for the high-dimensional quantile regression by smoothing the quantile loss, while \cite{cai2025statistical} proposed statistical inference procedures for high-dimensional rank regression by smoothing the rank loss. Nevertheless, inference for high-dimensional smoothed SVMs in the offline setting remains largely unexplored.
 
In the context of streaming datasets, moving beyond point estimation, recent attention has shifted towards real-time statistical inference for high-dimensional models. For high-dimensional linear models, \cite{han2024online} developed an online debiased Lasso framework, updating the summary statistics to correct shrinkage bias for statistical inference. Furthermore, \cite{luo2023online} proposed a debiased estimator for high-dimensional generalized linear models in streaming data, which employed approximate summary statistics to construct bias-corrected updates, ensuring valid statistical inference.
Despite these advances, establishing inference for models with non-smoothed loss functions remains computationally inefficient and challenging for ensuring valid statistical inference in the online setting. For example, although \cite{zhang2025renewable} provided an online debiased estimator for inference in high-dimensional SVMs based on the non-smoothed hinge loss, the use of the non-smoothed hinge loss deteriorates both computational efficiency and inferential validity, as illustrated in Section \ref{simulation}.
Building upon the success of offline smoothed methods, recent studies have extended convolution smoothing techniques to online high-dimensional quantile regression \cite[]{xie2025statistical} and convoluted rank regression \cite[]{WANG_rank_online}. 
Coupling the smoothed losses with approximate summary statistics, they derived debiased estimators to construct valid inference with improved computation efficiency. However, statistical inference for high-dimensional smoothed SVMs in the online setting has not been explored.

\subsection{Organization of the paper}
The rest of this paper is organized as follows. In Section \ref{offline method}, we develop an offline debiased Lasso estimator and provide a method to construct its confidence interval. In Section \ref{online method}, we incorporate the online updating method to construct an online debiased Lasso estimator and its confidence interval. In Section \ref{theory}, we report the theoretical properties for our offline and online methods. Simulation studies are provided in Section \ref{simulation} to evaluate the performance of our proposed offline and online smoothed debiased estimators. In Section \ref{application}, we apply the proposed method to the offline movie review dataset and the online 10-K corpus dataset. Discussions are given in Section \ref{discussion}. Additional simulation results can be found in Appendix \ref{simulation appendix}. 

\subsection{Notations} For vectors $\boldsymbol{u} = (u_1, u_2, \ldots,u_k)^T$ and $\boldsymbol{v} = (v_1, v_2, \ldots, v_k)^T$,  the inner product $\langle \boldsymbol{u},\boldsymbol{v} \rangle  = \boldsymbol{u}^T \boldsymbol{v} = \sum_{i = 1}^k u_i v_i$ , the $l_q$ norm $\Vert \boldsymbol{u} \Vert_q = (\sum_{i = 1}^k |u_i|^q)^{1/q}$ , and the $l_{\infty}$-norm $\Vert \boldsymbol{u} \Vert_{\infty} = \mathrm{max}_{1 \leq i \leq  k}\: u_i$. For any matrix $\boldsymbol{A} = \{a_{ij}\}_{i,j}^k \in \mathbb{R}^{k \times k}$, write $\Vert \boldsymbol{A} \Vert_{L_1} = \max_i\sum_{j=1}^k|a_{ij}|$ and $\Vert \boldsymbol{A} \Vert_{\infty} = \max_{i,j}|a_{ij}|$. For any $q \in [0, 1)$, define the off-diagonal row-wise $L_q$ norm as $\Vert \boldsymbol{A} \Vert_{L_q} = \max_{1 \leq i \leq k} \sum_{j = 1}^k h_q(a_{ij})$, where $h_q(x) = |x|^q$ if $q \in (0, 1)$ and $h_q(x) = I(x \neq 0)$ if $q = 0$. Let $\{\gamma_i(\boldsymbol{A})\}_{i=1}^{k}$ be the eigenvalue of the matrix $\boldsymbol{A}$, where $\gamma_1(\boldsymbol{A}) \geqslant \gamma_2(\boldsymbol{A}) \geqslant \ldots \geqslant\gamma_{k}(\boldsymbol{A})$. For two sequences $a_n$ and $b_n$, write $a_n = O(b_n)$ or $a_n \lesssim b_n$ if $|a_n| \leq C|b_n|$ for some constant $C>0$ independent of $n$, $a_n = o(b_n)$ if $\lim_{n\rightarrow\infty} a_n/b_n = 0$, and $a_n \asymp b_n$ if $a_n = O(b_n)$ and $b_n = O(a_n)$ hold. For two sequences of random variables $\{X_n\}$ and $\{Y_n\}$,  write $X_n = O_p(Y_n)$ if for every $\epsilon > 0$, there exists a constant $M_\epsilon > 0$ such that $ \sup_{n \geq 1} P(|X_n| > M_\epsilon |Y_n|) \leq \epsilon$, 
and write $X_n = o_p(Y_n)$ if for every ${\varepsilon} > 0$, $\lim_{n \to \infty} P(|X_n| \geq {\varepsilon} |Y_n|) = 0$.

\section{Methodology\label{Methodology}}
\subsection{Smoothed SVM for the offline dataset\label{offline method}}
Given high-dimensional features $\boldsymbol{X}\in \mathbb{R}^p$ and response $Y \in \{-1, 1\}$, our primary interest is to build a classifier that predicts the response \(Y\) from the features \(\boldsymbol{X}\). Define $\tilde{\boldsymbol{X}} = (1, \boldsymbol{X}^\top)^\top$. A widely used approach for classification is the SVM \cite[]{koo2008bahadur}, which seeks a classifier \(\mathrm{sign}(\tilde{\boldsymbol{X}}^\top \boldsymbol{\beta}^*)\) for prediction, where the parameter \(\boldsymbol{\beta}^*\) is assumed \(s\)-sparse ($ \|\boldsymbol{\beta}^*\|_0=s$) as feature is high dimensional. Here, the parameter \(\boldsymbol{\beta}^*\) is defined by:
\begin{equation}
    \boldsymbol{\beta}^* =\underset{\boldsymbol{\beta} \in \mathbb{R}^{p + 1}}{\mathrm{arg\:min}}\: E\{l(Y \tilde{\boldsymbol{X}}^\top \boldsymbol{\beta})\},
    \label{originial hinge loss}
\end{equation}
where $l(u)=\mathrm{max}\{1 - u, 0\}, u\in\mathbb{R}$ is the hinge loss.

Consider the offline dataset \(\tilde{D}_{\mathrm{off}}=\{Y_i,\boldsymbol{X}_i\}_{i=1}^n\).
A popular method to estimate the sparse parameter \(\boldsymbol{\beta}^*\) is through minimizing the empirical hinge loss with a Lasso penalty \cite[]{peng2016error, lian2018divide}: 
\begin{equation}
     \underset{\boldsymbol{\beta} \in \mathbb{R}^{p + 1}}{\mathrm{arg\:min}}\: \frac{1}{n}\sum_{i=1}^nl(Y_i \tilde{\boldsymbol{X}_i}^\top \boldsymbol{\beta}) + \lambda \|\boldsymbol{\beta}_{-0}\|_1.
    \label{original hinge loss}
\end{equation}

Since the hinge loss \(l(\cdot)\) is not differentiable, it is challenging to solve the optimization problem (\ref{original hinge loss}) and to derive a confidence bound based on the solution of the optimization. Specifically, first, the non-differentiability poses significant computational burdens to solve the optimization as fast high-dimensional optimization algorithms, such as the proximal gradient descent, are only feasible for differentiable loss \cite[]{FISTA, parikh2014proximal}. Second, the asymptotic confidence interval for $\boldsymbol{\beta}^*$ based on the hinge loss (\ref{original hinge loss}) may not perform well in finite samples due to the non-differentiability of the hinge loss, as demonstrated in \cite{Rybak_jmlr} and in our simulations in Section \ref{simulation}. 
Therefore, we propose to smooth the hinge loss function via the kernel smoother \cite[]{fernandes2021qrsmoothing, density-convoluted}. 

We begin by introducing some notations. Let $K(\cdot)$ denote a kernel function (e.g., Gaussian or uniform kernel), with its cumulative distribution function (CDF) defined as $\mathcal{K}(u) = \int_{-\infty}^u K(t) dt$. For a bandwidth parameter $h > 0$, the scaled kernel function is given by $K_h(u) = h^{-1}K(u/h)$, and its corresponding CDF is $\mathcal{K}_h(u) = \mathcal{K}(u/h)$. 

Note that the hinge loss in (\ref{original hinge loss}) can be expressed as
\begin{align} \label{eq:intuition}
    \frac{1}{n}\sum_{i=1}^nl(Y_i \tilde{\boldsymbol{X}_i}^\top \boldsymbol{\beta}) = \int_{-\infty}^\infty l(t) d \widehat{F}(t; \boldsymbol{\beta}),
\end{align}
where $\widehat{F}(t; \boldsymbol{\beta}) = n^{-1} \sum_{i=1}^n I(Y_i \tilde{\boldsymbol{X}}_i^\top \boldsymbol{\beta} \leq t)$ is the empirical CDF. The non-differentiability of \(\widehat{F}(t; \boldsymbol{\beta})\) with respect to \(\boldsymbol{\beta}\) causes the loss function (\ref{eq:intuition}) to be non-differentiable with respect to \(\boldsymbol{\beta}\). Therefore, we propose to replace the the empirical CDF \(\widehat{F}(t; \boldsymbol{\beta})\) by a differentiable kernel CDF:

$$
\widehat{F}_h(t;\boldsymbol{\beta}) =  \frac{1}{n} \sum_{i = 1}^n \mathcal{K}_h(t - Y_i \tilde{\boldsymbol{X}}_i^\top \boldsymbol{\beta}).
$$

By replacing the non-smoothed empirical CDF with this smoothed CDF, the empirical hinge loss (\ref{eq:intuition}) becomes the empirical smoothed hinge loss:
$$
\int_{-\infty}^\infty l(t) d \widehat{F}_h(t;\boldsymbol{\beta}) =  \frac{1}{n} \sum_{i = 1}^n \int_{-\infty}^\infty l(t) K_h(t - Y_i \tilde{\boldsymbol{X}}_i^\top \boldsymbol{\beta}) d t = \frac{1}{n} \sum_{i=1}^n l_h(Y_i \tilde{\boldsymbol{X}}_i^\top \boldsymbol{\beta}),
$$
where $l_h(u) = (l * K_h)(u) = \int_{-\infty}^\infty l(t) K_h(t - u) dt$ is the convolution-smoothed hinge loss function. In this paper, we choose the Gaussian kernel \(K(t)=\{2\pi\}^{-1/2} \exp(-t^2/2)\) for implementation as it is a common choice in the literature and it works well in our simulation and application. When the Gaussian kernel is employed, the smoothed hinge loss is
\begin{align*}
    l_h(u) = \frac{h}{\sqrt{2\pi}} \exp\left(-\frac{(1-u)^2}{2h^2}\right) + (1-u) \int_{-\infty}^{\frac{1-u}{h}} \frac{1}{\sqrt{2\pi}} \exp\left(-\frac{t^2}{2}\right)dt.
\end{align*} 

\begin{figure}[htbp]
    \centering
    \includegraphics[width=0.6\linewidth]{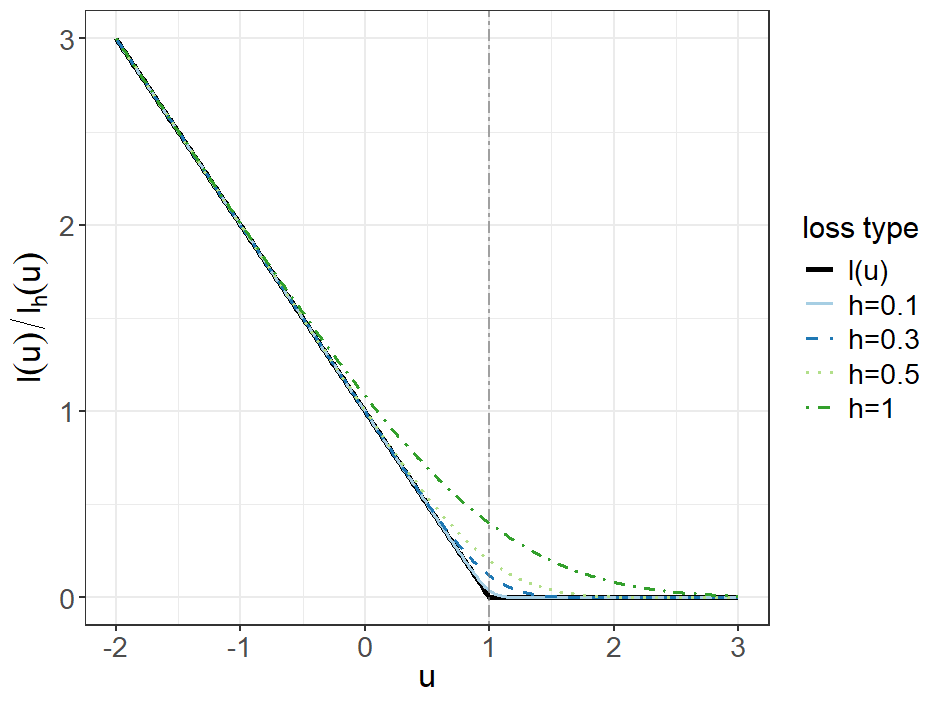}
    \caption{A comparison between the hinge loss \(l(u)\) and its Gaussian-smoothed variant \(l_h(u)\). The black bold curve represents \(l(u)\), while the remaining curves correspond to \(l_h(u)\) for different values of \(h\); for instance, \(h = 0.1\) denotes \(l_{0.1}(u)\).
}
    \label{fig:smooth comparison}
\end{figure}

Figure~\ref{fig:smooth comparison} depicts the hinge loss \(l(u)\) alongside its Gaussian-smoothed counterpart \(l_h(u)\) for various choices of the bandwidth \(h\). It is clear that, as \(h\) decreases, the smoothed loss \(l_h(u)\) converges to the original hinge loss \(l(u)\).

By combining the smoothed hinge loss with the Lasso penalty, we propose to estimate the \(\boldsymbol{\beta}^*\) by:

\begin{equation}
  \widehat{\boldsymbol{\beta}}^{\mathrm{off}} = \underset{\boldsymbol{\beta} \in \mathbb{R}^{p + 1}}{\mathrm{arg\:min}}\: \bigg\{ \widehat{L}({\boldsymbol{\beta}}, \tilde{D}_{\mathrm{off}}, h)    + \lambda \Vert \boldsymbol{\beta}_{-0} \Vert_1 \bigg\},
  \label{offline smooth lasso estimator}
\end{equation}
where $\widehat{L}({\boldsymbol{\beta}}, \tilde{D}_{\mathrm{off}}, h) = n^{-1} \sum_{i=1}^n l_h(Y_i \tilde{\boldsymbol{X}}_i^\top \boldsymbol{\beta})$. There are two hyperparameters, \(h\) and \(\lambda\), that need to be specified. For the choice of the bandwidth parameter \(h\), we recommend $h = \{5 \log (p)/n\}^{1/4}$, which performs well in our simulation and satisfies the conditions to establish consistency in Theorem \ref{theorem: offline beta consistency}. The tuning parameter \(\lambda\) can be chosen by 5-fold cross-validation. 

Overall, with kernel smoothing, our objective function (\ref{offline smooth lasso estimator}) is now a combination of a differentiable loss function and the Lasso penalty. Therefore, an efficient optimization algorithm and the debiased Lasso technique for constructing the confidence bound can be utilized. In this paper, we choose the proximal gradient descent algorithm for efficient computation. More details about this algorithm can be found in \cite{proximal}.

We now explicitly detail how to construct a confidence bound for \(\boldsymbol{\beta}^*\) using the debiased Lasso technique. The debiased Lasso technique works by integrating the gradient and Hessian of the smoothed hinge loss to debias the Lasso estimates \(\widehat{\boldsymbol{\beta}}^{\mathrm{off}}\), and then estimating the variance of the debiased Lasso estimates to construct an asymptotically normal confidence bound.

Specifically, according to the theory in \cite{debias_lasso_vande}, we develop a debiased Lasso estimate:
\begin{equation}
\label{offline oracle debias estimator}
\widehat{\boldsymbol{\beta}}^{\mathrm{off}}_j - (\widehat{\boldsymbol{H}}^{-1})_j^\top \left\{ \nabla \widehat{L}(\widehat{\boldsymbol{\beta}}^{\mathrm{off}}, \tilde{D}_{\mathrm{off}}, h) \right\},
\end{equation}
where $
    \nabla \widehat{L}(\widehat{\boldsymbol{\beta}}^{\mathrm{off}}, \tilde{D}_{\mathrm{off}}, h)= n^{-1}\sum_{i = 1}^{n} \{-\mathcal{K}_h(1-Y_i\tilde{\boldsymbol{X}}_i^T \widehat{\boldsymbol{\beta}}^{\mathrm{off}})Y_i\tilde{\boldsymbol{X}}_i \}$ represents the gradient of $\widehat{L}(\boldsymbol{\beta}, \tilde{D}_{\mathrm{off}}, h)$ with respect to \(\boldsymbol{\beta}=\widehat{\boldsymbol{\beta}}^{\mathrm{off}}\), $\widehat{\boldsymbol{H}} = \nabla^2 \widehat{L}(\widehat{\boldsymbol{\beta}}^{\mathrm{off}}, \tilde{D}_{\mathrm{off}}, h)= n^{-1}\sum_{i = 1}^{n} K_h(1-Y_i\tilde{\boldsymbol{X}}_i^T \widehat{\boldsymbol{\beta}}^{\mathrm{off}})\allowbreak \tilde{\boldsymbol{X}}_i\tilde{\boldsymbol{X}}_i^T
$ is the Hessian matrix, and $(\widehat{\boldsymbol{H}}^{-1})_j$ is the $j$-th column of $\widehat{\boldsymbol{H}}^{-1}$. 

In high-dimensional settings, the Hessian matrix $\widehat{\boldsymbol{H}}$ is frequently singular and hence non-invertible, which makes the estimate (\ref{offline oracle debias estimator}) questionable. Therefore, we further propose to estimate the inverse of the Hessian matrix using the CLIME method \cite[]{clime} as follows:
\begin{equation}
    \widehat{\boldsymbol{\Theta}}_j = \underset{\boldsymbol{\Theta}_j \in \mathbb{R}^{p+1}}{\text{arg min}}\Vert \boldsymbol{\Theta}_j \Vert _1, \:\text{s.t.} \big\Vert \widehat{\boldsymbol{H}} \boldsymbol{\Theta}_j  - \boldsymbol{e}_j \big\Vert _{\infty} \leqslant \delta, j=1,...,p,
    \label{offline information estimation}
\end{equation}\
where \(\widehat{\boldsymbol{\Theta}}_j\) can approximate \((\widehat{\boldsymbol{H}}^{-1})_j\), $\boldsymbol{e}_j$ denotes the j-th canonical basis vector in $\mathbb{R}^{p+1}$ (i.e., its j-th element equals 1 while others are 0), and $\delta$ is a tuning parameter can be chosen via 5-fold cross-validation. This optimization problem can be solved via linear programming. For further details, we refer to \cite{clime}.

Our final debiased Lasso estimate is
\begin{equation}
\widehat{\boldsymbol{\beta}}^{\mathrm{off},\mathrm{de}}_j = \widehat{\boldsymbol{\beta}}_j^{\mathrm{off}} - (\widehat{\boldsymbol{\Theta}}_{j})^{\top} \bigg\{ \nabla \widehat{L}(\widehat{\boldsymbol{\beta}}^{\mathrm{off}}, \tilde{D}_{\mathrm{off}}, h)\bigg\}, \quad j = 1,\ldots,p.
\label{offline debias estimator}
\end{equation}

We have established the asymptotic normality of the debiased Lasso estimate in Theorem \ref{theorem: offline asymptotic}. This theorem can be utilized to construct a 95\% Wald-type confidence bound. Specifically, we can first estimate the variance of (\ref{offline debias estimator}) by $\widehat{{\sigma}}_{j}^2 = (\widehat{\boldsymbol{\Theta}}_j)^\top \widehat{\boldsymbol{\boldsymbol{\Sigma}}} \widehat{\boldsymbol{\Theta}}_j$, where 
\begin{equation}
\label{sigma: offline sigma}
    \widehat{\boldsymbol{\boldsymbol{\Sigma}}} = \frac{1}{n}\sum_{i=1}^{n} \nabla l_h(Y_i \tilde{\boldsymbol{X}}_i^\top \widehat{\boldsymbol{\beta}}^{\mathrm{off}})
\big\{\nabla l_h(Y_i \tilde{\boldsymbol{X}}_i^\top \widehat{\boldsymbol{\beta}}^{\mathrm{off}})\big\}^\top = \frac{1}{n} \sum_{i = 1}^{n} \tilde{\boldsymbol{X}}_{i} \tilde{\boldsymbol{X}}_{i}^\top \{\mathcal{K}_{h}(1-Y_{i}\tilde{\boldsymbol{X}}_{i}^\top \widehat{\boldsymbol{\beta}}^{\mathrm{off}})\}^2.
\end{equation}

The 95\% Wald-type asymptotic normal confidence interval then is:
\begin{equation}
\label{offline interval}
  \left(\widehat{\boldsymbol{\beta}}^{\mathrm{off},\mathrm{de}}_j - 1.96 \cdot\widehat{{\sigma}}_{j} / \sqrt{n}, \widehat{\boldsymbol{\beta}}^{\mathrm{off},\mathrm{de}}_j  + 1.96 \cdot\widehat{{\sigma}}_{j} / \sqrt{n}\right).  
\end{equation}

We have summarized our offline SVM algorithm in Algorithm \ref{algorithm_offline}. 

\begin{algorithm}
\caption{Offline debiased Lasso algorithm.}
\label{algorithm_offline}
\begin{algorithmic}[1]
\renewcommand{\algorithmicrequire}{\textbf{Input:}}
\renewcommand{\algorithmicensure}{\textbf{Output:}}
\REQUIRE Offline dataset $\tilde{D}_{\mathrm{off}}=\{Y_i,\tilde{\boldsymbol{X}}_i\}_{i=1}^n$, smoothing bandwidth $h$, hyperparameters sequence of $\lambda$ and $\delta$;
\ENSURE Debiased offline estimate for the parameter of the linear classifier, \(\boldsymbol{\beta}^*\), and its 95\% confidence interval.
\STATE Calculate the Lasso estimate:  $\widehat{\boldsymbol{\beta}}^{\mathrm{off}} = {\arg\,\min}_{\boldsymbol{\beta} \in \mathbb{R}^{p + 1}}\: \big\{ n^{-1} \sum_{i=1}^n l_h(Y_i \tilde{\boldsymbol{X}}_i^\top \boldsymbol{\beta}) + \lambda \Vert \boldsymbol{\beta}_{-0} \Vert_1 \big\}$, where \(\lambda\) is tuned by 5-fold cross-validation.
\STATE Compute the gradient and the Hessian matrix of the smoothed hinge loss:
$
    \nabla \widehat{L}(\widehat{\boldsymbol{\beta}}^{\mathrm{off}}, \tilde{D}_{\mathrm{off}}, h)= n^{-1}\sum_{i = 1}^{n} \{-\mathcal{K}_h(1-Y_i\tilde{\boldsymbol{X}}_i^T \widehat{\boldsymbol{\beta}}^{\mathrm{off}})Y_i\tilde{\boldsymbol{X}}_i \}$, and $
\widehat{\boldsymbol{H}} = \nabla^2 \widehat{L}(\widehat{\boldsymbol{\beta}}^{\mathrm{off}}, \tilde{D}_{\mathrm{off}}, h) = n^{-1}\sum_{i = 1}^{n} K_h(1-Y_i\tilde{\boldsymbol{X}}_i^T \widehat{\boldsymbol{\beta}}^{\mathrm{off}})\tilde{\boldsymbol{X}}_i\tilde{\boldsymbol{X}}_i^T.
$
\STATE Estimate the precision matrix \(\widehat{\boldsymbol{H}}^{-1}\) by \(\widehat{\boldsymbol{\Theta}}\), where the \(j\)-th column of \(\widehat{\boldsymbol{\Theta}}\) is defined by
$\widehat{\boldsymbol{\Theta}}_{j} = {\arg\,\min}_{\boldsymbol{\Theta}_j \in \mathbb{R}^{p+1}}\Vert \boldsymbol{\Theta}_j \Vert _1, \:\text{s.t.} \big\Vert \widehat{\boldsymbol{H}} \boldsymbol{\Theta}_j  - \boldsymbol{e}_j \big\Vert _{\infty} \leqslant \delta,$ and $\delta$ is tuned by 5-fold cross-validation.
\STATE Calculate the debiased estimate $\widehat{\boldsymbol{\beta}}^{\mathrm{off},\mathrm{de}}_j = \widehat{\boldsymbol{\beta}}_j^{\mathrm{off}} - \widehat{\boldsymbol{\Theta}}_{j}^{\top} \big\{ \nabla \widehat{L}(\widehat{\boldsymbol{\beta}}^{\mathrm{off}}, \tilde{D}_{\mathrm{off}}, h)\big\}$.
\STATE Estimate the variance of the debiased estimate $\widehat{\boldsymbol{\beta}}^{\mathrm{off},\mathrm{de}}_j$ by $\widehat{{\sigma}}_{j}^2 = (\widehat{\boldsymbol{\Theta}}_{j})^\top \widehat{\boldsymbol{\boldsymbol{\Sigma}}} \widehat{\boldsymbol{\Theta}}_{j}$, where $\widehat{\boldsymbol{\boldsymbol{\Sigma}}} = n^{-1}\sum_{i = 1}^{n} \tilde{\boldsymbol{X}}_{i} \tilde{\boldsymbol{X}}_{i}^\top \{\mathcal{K}_{h}(1-Y_{i}\tilde{\boldsymbol{X}}_{i}^\top \widehat{\boldsymbol{\beta}}^{\mathrm{off}})\}^2$. 
\STATE Construct 95\% confidence interval:
$\big(\widehat{\boldsymbol{\beta}}^{\mathrm{off},\mathrm{de}}_j - 1.96 \cdot\widehat{{\sigma}}_{j} / \sqrt{n}, \widehat{\boldsymbol{\beta}}^{\mathrm{off},\mathrm{de}}_j  + 1.96 \cdot\widehat{{\sigma}}_{j} / \sqrt{n}\big).$
\STATE Output the debiased Lasso estimate \(\widehat{\boldsymbol{\beta}}^{\mathrm{off},\mathrm{de}}_j\) and its 95\% confidence interval.
\end{algorithmic}
\end{algorithm}

\subsection{Smoothed SVM for the online streaming data\label{online method}}
In this subsection, we discuss how to construct the smoothed SVM when the data arrives in batches.
Let $D_b = \{Y_{bi}, \boldsymbol{X}_{bi}\}_{i=1}^{n_b}$ represent the observations for the $b$-th batch ($b = 1,\ldots,B$). The main difficulty with streaming data is that we can not store the individual observations of the previous batches but only some summary statistics due to the storage limit. This requires us to develop an updating scheme for the smoothed SVM that relies solely on the current batch's data together with the summary statistics retained from prior batches.

We now detail the construction of the online smoothed SVM. We begin by introducing the oracle smoothed SVM, which assumes access to all individual observations from previous batches. We then discuss how to approximate this oracle using only the summary statistics retained from earlier batches, thereby obtaining our online smoothed SVM.

Following the offline smoothed SVM (\ref{offline smooth lasso estimator}), when all individual observations from previous batches are available, an oracle estimator for the data up to the $b$-th batch is:
\begin{equation}
  \widehat{\boldsymbol{\beta}}_b^o = \underset{\boldsymbol{\beta} \in \mathbb{R}^{p + 1}}{\mathrm{arg\:min}}\: \bigg\{ \frac{1}{N_b} \sum_{t=1}^b n_t\widehat{L}(\boldsymbol{\beta}, D_t, h_t) + \lambda_b \Vert \boldsymbol{\beta}_{-0} \Vert_1 \bigg\},
  \label{online oracle complete data estimator}
\end{equation}
where $N_b = \sum_{t=1}^b n_t$ is the cumulative sample size, $\widehat{L}(\boldsymbol{\beta}, D_t, h_t)=n_t^{-1}\sum_{i=1}^{n_t}l_{h_t}(Y_{ti}\tilde{\boldsymbol{X}}_{ti}^\top \boldsymbol{\beta})$ is the loss for \(t\)-th batch, and $h_t$ denotes the bandwidth for the $t$-th batch. 

Since individual observations from earlier batches cannot be retained due to storage limitations, the oracle smoothed SVM (\ref{online oracle complete data estimator}) is not feasible for streaming data. In particular, the term \(N_{b}^{-1}\sum_{t=1}^{b-1} n_t \widehat{L}(\boldsymbol{\beta}, D_t, h_t)\) in (\ref{online oracle complete data estimator}) cannot be evaluated when the individual data from prior dataset, \(D_t, t=1,..,b-1\), are unavailable. Therefore, we propose to approximate the term \(N_{b}^{-1}\sum_{t=1}^{b-1} n_t \widehat{L}(\boldsymbol{\beta}, D_t, h_t)\) using summary statistics. 

In particular, we first discuss how to approximate \(N_{b}^{-1}\sum_{t=1}^{b-1} n_t \widehat{L}(\boldsymbol{\beta}, D_t, h_t)\) using summary statistics when \(b=2\). Note that the oracle estimate \(\widehat{\boldsymbol{\beta}}_1^o\) is feasible as no prior data batch should be incorporated for the estimate \(\widehat{\boldsymbol{\beta}}_1^o\). For clear presentation, denote \(\widehat{\boldsymbol{\beta}}_1^{\mathrm{on}}=\widehat{\boldsymbol{\beta}}_1^o\). When \(b=2\), applying Taylor expansion for the term needed to be approximated, \(\frac{n_1}{N_2}\widehat{L}(\boldsymbol{\beta}, D_2, h_2)\), with respect to \(\boldsymbol{\beta}=\widehat{\boldsymbol{\beta}}_1^{\mathrm{on}}\), we obtain: 
\begin{align}
   \notag\frac{n_1}{N_2} \widehat{L}(\boldsymbol{\beta}, D_t, h_t) 
    &= \underbrace{\frac{n_1}{N_2} \widehat{L}(\widehat{\boldsymbol{\beta}}_{1}^{\mathrm{on}}, D_1, h_1)}_{U_1}  + (\boldsymbol{\beta} - \widehat{\boldsymbol{\beta}}_{1}^{\mathrm{on}})^{\top}\underbrace{\bigg\{\frac{n_1}{N_2} \nabla \widehat{L}(\widehat{\boldsymbol{\beta}}_{1}^{\mathrm{on}}, D_1, h_1) \bigg\}}_{U_2} \\
   \notag &+ \frac{1}{2} (\boldsymbol{\beta} - \widehat{\boldsymbol{\beta}}_{1}^{\mathrm{on}})^\top\underbrace{\bigg\{\frac{n_1}{N_2} \nabla^2 \widehat{L}(\widehat{\boldsymbol{\beta}}_{1}^{\mathrm{on}}, D_1, h_1)\bigg\}}_{U_3}(\boldsymbol{\beta} - \widehat{\boldsymbol{\beta}}_{1}^{\mathrm{on}}) \\
   \label{eq:tayler} &+ \underbrace{o_p\big(\|\boldsymbol{\beta} - \widehat{\boldsymbol{\beta}}_{1}^{\mathrm{on}}\|^2\big)}_{R}.
\end{align}

Since all terms related to \(\boldsymbol{\beta}\) on the right hand of the equation (\ref{eq:tayler}), can be calculated by summary statistics $\widehat{\boldsymbol{\beta}}_{1}^{\mathrm{on}}$, $\nabla \widehat{L}(\widehat{\boldsymbol{\beta}}_{1}^{\mathrm{on}}, D_1, h_1)$ and $\nabla^2 \widehat{L}(\widehat{\boldsymbol{\beta}}_{1}^{\mathrm{on}}, D_1, h_1)$, this approximation can be utilized in the oracle objective function (\ref{online oracle complete data estimator}) to construct an approximation of the oracle estimate \(\widehat{\boldsymbol{\beta}}_2^o\).

To be clarified, the first term $U_1$ of (\ref{eq:tayler}) is a constant with respect to $\boldsymbol{\beta}$, and therefore can be omitted when approximating the oracle objective function (\ref{online oracle complete data estimator}). The $U_2$ in the second term of (\ref{eq:tayler}) is asymptotically ignorable, since $\| U_2 \|_{\infty} = O_p(\lambda_{1})=o_p(1)$ according to the Karush-Kuhn-Tucker (KKT) conditions of the Lasso estimate \(\widehat{\boldsymbol{\beta}}_1^{\mathrm{on}}\). Therefore, we can ignore the second term when approximating the oracle optimization objective function. The third term of (\ref{eq:tayler}) should be retained as $U_3$ in the third term is not asymptotically ignorable. The fourth term \(R\) is a higher order of the third term and therefore can be omitted.

Overall, we approximate the oracle estimate \(\widehat{\boldsymbol{\beta}}_2^o\) by \(\widehat{\boldsymbol{\beta}}_2^{\mathrm{on}}\) as follows:
$$
\widehat{\boldsymbol{\beta}}_2^{\mathrm{on}} = \underset{\boldsymbol{\beta} \in \mathbb{R}^{p + 1}}{\mathrm{arg\:min}}\: \bigg[ \frac{1}{2} (\boldsymbol{\beta} - \widehat{\boldsymbol{\beta}}_{1}^{\mathrm{on}})^\top\bigg\{\frac{n_1}{N_2} \nabla^2 \widehat{L}(\widehat{\boldsymbol{\beta}}_{1}^{\mathrm{on}}, D_1, h_1)\bigg\}(\boldsymbol{\beta} - \widehat{\boldsymbol{\beta}}_{1}^{\mathrm{on}}) + \frac{n_2}{N_2} \widehat{L}(\boldsymbol{\beta}, D_2, h_2) + \lambda_2 \Vert \boldsymbol{\beta}_{-0} \Vert_1 \bigg].
$$

Following this idea, we can obtain the following approximation for any \(b\): 
\begin{equation}
 \widehat{\boldsymbol{\beta}}_b^{\mathrm{on}} = \underset{\boldsymbol{\beta} \in \mathbb{R}^{p + 1}}{\mathrm{arg\:min}}\: \bigg\{\tilde{L}_b(\boldsymbol{\beta}, h_b) + \lambda_b \Vert \boldsymbol{\beta}_{-0} \Vert_1 \bigg\}, b=1,...,B,
    \label{online lasso bias estimator}
\end{equation}
where $\tilde{L}_b(\boldsymbol{\beta}, h_b) = \frac{1}{N_b}\big\{\frac{N_{b-1}}{2}(\boldsymbol{\beta} - \widehat{\boldsymbol{\beta}}_{b-1}^{\mathrm{on}})^\top \widehat{\boldsymbol{H}}_{b - 1} (\boldsymbol{\beta} - \widehat{\boldsymbol{\beta}}_{b-1}^{\mathrm{on}}) + n_b\widehat{L}(\boldsymbol{\beta}, D_b, h_b)\big\}$, and $\widehat{\boldsymbol{H}}_{b-1} = N_{b-1}^{-1}\sum_{t=1}^{b-1} n_t \nabla^2 \widehat{L}(\widehat{\boldsymbol{\beta}}_{t}^{\mathrm{on}}, D_t, h_t)$ is a summary statistics.

This optimization problem can be solved by proximal gradient descent. For the hyperparameters, \(h_b\) and \(\lambda_b\). We recommend $h_b = (5 \log (p)/N_b)^{1/4}$ as it performs well in our simulation and ensures the consistent result established in Theorem \ref{theorem:online bound}. The parameter 
\(\lambda_b\) is selected via 5-fold cross-validation. 

We now discuss how to construct a confidence interval for \(\boldsymbol{\beta}^*\) based on the online Lasso estimate (\ref{online lasso bias estimator}). Similar to the offline case discussed in Section \ref{offline method}, we begin by removing the bias of the online Lasso estimate (\ref{online lasso bias estimator}) by the debiased Lasso technique and then estimate the variance of the debiased Lasso estimate to construct the asymptotically normal confidence bound.

We propose the following online debiased estimator, inspired by the offline debiased framework in (\ref{offline debias estimator}) and adapted through suitable approximations using summary statistics to ensure its feasibility in the online setting:
\begin{align}
\label{online debias estimator}
     \widehat{\boldsymbol{\beta}}_{b,j}^{\mathrm{on},\mathrm{de}} &= \widehat{\boldsymbol{\beta}}_{b,j}^{\mathrm{on}} - \frac{1}{N_b}  (\widehat{\boldsymbol{\Theta}}_{bj})^{\top} \sum_{t = 1}^b n_t \nabla \widehat{L}(\widehat{\boldsymbol{\beta}}_t^{\mathrm{on}}, D_t, h_t)\\ \notag
     &\quad + \frac{1}{N_b} (\widehat{\boldsymbol{\Theta}}_{bj})^{\top} \sum_{t = 1}^{b} n_t \nabla^2 \widehat{L}(\widehat{\boldsymbol{\beta}}_{t}^{\mathrm{on}}, D_t, h_t)(\widehat{\boldsymbol{\beta}}_t^{\mathrm{on}} - \widehat{\boldsymbol{\beta}}_b^{\mathrm{on}}),
\end{align}
where \(\widehat{\boldsymbol{\Theta}}_{bj}\) is an approximation of the $j$-th column of the inverse of the matrix $\widehat{\boldsymbol{H}}_b$, defined by
\begin{equation}
    \widehat{\boldsymbol{\Theta}}_{bj} = \underset{\boldsymbol{\Theta}_{bj} \in \mathbb{R}^{p+1}}{\text{arg min}}\Vert \boldsymbol{\Theta}_{bj} \Vert _1, \:\text{s.t.} \big\Vert \widehat{\boldsymbol{H}}_b \boldsymbol{\Theta}_{bj} - \boldsymbol{e}_j \big\Vert _{\infty} \leqslant \delta_{b}.
    \label{online information estimation}
\end{equation}

Here, $\delta_b$ is a tuning parameter for the $b$-th batch, which can be chosen via 5-fold cross-validation. The optimization problem (\ref{online information estimation}) can be solved by linear programming \cite[]{clime}. 

The online debiased Lasso estimate (\ref{online debias estimator}) includes two debiased terms. The first debiased term is analogous to the debiased term in the offline setting (\ref{offline debias estimator}), serving to eliminate the bias induced by the Lasso penalty. The second debiased term is distinctive to the online setting, introduced to remove the bias incurred by approximations using summary statistics.
We have established the asymptotic normality of the debiased Lasso estimate (\ref{online debias estimator}) in Theorem \ref{theorem:debias theory}. This result can be leveraged to construct a 95\% Wald-type confidence interval.
Specifically, we can first estimate the variance of (\ref{online debias estimator}) by $\widehat{{\sigma}}_{bj}^2 = \widehat{\boldsymbol{\Theta}}_{bj}^\top \widehat{\boldsymbol{\boldsymbol{\Sigma}}}_b \widehat{\boldsymbol{\Theta}}_{bj}$, where the summary statistics 
\begin{equation}
\label{sigma: online sigma}
    \widehat{\boldsymbol{\boldsymbol{\Sigma}}}_b = N_b^{-1} \sum_{t = 1}^b \sum_{i = 1}^{n_t} \tilde{\boldsymbol{X}}_{ti} \tilde{\boldsymbol{X}}_{ti}^\top \{\mathcal{K}_{h_t}(1-Y_{ti}\tilde{\boldsymbol{X}}_{ti}^\top \widehat{\boldsymbol{\beta}}_t^{\mathrm{on}})\}^2.
\end{equation}

Then, we can construct the 95\% confidence interval as follows: 
\begin{equation}
\label{online interval}
    \left(\widehat{\boldsymbol{\beta}}_{b,j}^{\mathrm{on},\mathrm{de}} - 1.96 \cdot\widehat{{\sigma}}_{bj} / \sqrt{N_b}, \widehat{\boldsymbol{\beta}}_{b,j}^{\mathrm{on},\mathrm{de}}  + 1.96 \cdot\widehat{{\sigma}}_{bj} / \sqrt{N_b}\right).
\end{equation}

We have summarized the online algorithm in Algorithm \ref{algorithm_online}. Here, we define the summary statistics 
\begin{align}
\label{summary statistics}
    \boldsymbol{S}_{b,1} &= \sum_{t = 1}^b n_t \nabla \widehat{L}(\widehat{\boldsymbol{\beta}}_t^{\mathrm{on}}, D_t, h_t),\  \boldsymbol{S}_{b,2} = \sum_{t = 1}^b n_t \nabla^2 \widehat{L}(\widehat{\boldsymbol{\beta}}_t^{\mathrm{on}}, D_t, h_t) \widehat{\boldsymbol{\beta}}_t^{\mathrm{on}}.
\end{align}

We update the summary statistics in our algorithm, including \( \boldsymbol{S}_{b,1}\), \( \boldsymbol{S}_{b,2}\), \(\widehat{\boldsymbol{H}}_{b}\), and \(\widehat{\boldsymbol{\boldsymbol{\Sigma}}}_b\), as follows: 
\begin{align}
\label{summary statistics update case}
    \boldsymbol{S}_{b,1} &= \boldsymbol{S}_{b-1,1} + n_b \nabla \widehat{L}(\widehat{\boldsymbol{\beta}}_b^{\mathrm{on}}, D_b, h_b),\  \boldsymbol{S}_{b,2} = \boldsymbol{S}_{b-1,2} + n_b \nabla^2 \widehat{L}(\widehat{\boldsymbol{\beta}}_b^{\mathrm{on}}, D_b, h_b) \widehat{\boldsymbol{\beta}}_b^{\mathrm{on}}, \\ \notag
    \widehat{\boldsymbol{H}}_{b} &= \frac{1}{N_b}\Big\{N_{b-1} \widehat{\boldsymbol{H}}_{b-1} + n_b \nabla^2 \widehat{L}(\widehat{\boldsymbol{\beta}}_{b}^{\mathrm{on}}, D_b, h_b)\Big\}, \\ \notag \widehat{\boldsymbol{\boldsymbol{\Sigma}}}_b &= \frac{1}{N_b} \Big\{ N_{b-1} \widehat{\boldsymbol{ \boldsymbol{\Sigma}}}_{b-1} + \sum_{i = 1}^{n_b} \tilde{\boldsymbol{X}}_{bi} \tilde{\boldsymbol{X}}_{bi}^\top \{\mathcal{K}_{h_b}(1-Y_{bi}\tilde{\boldsymbol{X}}_{bi}^\top \widehat{\boldsymbol{\beta}}_b^{\mathrm{on}})\}^2 \Big\},
\end{align}

\begin{algorithm}[ht]
\caption{Online debiased Lasso algorithm}
\label{algorithm_online}
\begin{algorithmic}[1] 
\renewcommand{\algorithmicrequire}{\textbf{Input:}}
\renewcommand{\algorithmicensure}{\textbf{Output:}}
\REQUIRE For each $b = 1, \ldots, B$, online dataset $D_b = \{Y_{bi}, \boldsymbol{X}_{bi}\}_{i=1}^{n_b}$, smoothing bandwidth $h_b$, hyperparameter sequence of $\lambda_b$ and $\delta_b$;
\ENSURE Debiased online estimate for the parameter of the linear classifier, $\boldsymbol{\beta}^*$, and its 95\% confidence interval.

\STATE Initialize $\boldsymbol{S}_{0,1} = \boldsymbol{0}$, $\boldsymbol{S}_{0,2} = \boldsymbol{0}$, $\widehat{\boldsymbol{H}}_0 = \boldsymbol{0}$, $\widehat{{\boldsymbol{\Sigma}}}_{0} = \boldsymbol{0}$, and  $\widehat{\boldsymbol{\beta}}_0^{\mathrm{on}} = \boldsymbol{0}$,

\FOR{$b = 1, 2, \ldots, B$}
    \STATE 1: Update the online Lasso estimate:
    $$ \widehat{\boldsymbol{\beta}}_b^{\mathrm{on}} = \underset{\boldsymbol{\beta} \in \mathbb{R}^{p + 1}}{\mathrm{arg\:min}}\: \bigg\{\frac{1}{N_b}\big\{\frac{N_{b-1}}{2}(\boldsymbol{\beta} - \widehat{\boldsymbol{\beta}}_{b-1}^{\mathrm{on}})^\top \widehat{\boldsymbol{H}}_{b - 1} (\boldsymbol{\beta} - \widehat{\boldsymbol{\beta}}_{b-1}^{\mathrm{on}}) + n_b\widehat{L}(\boldsymbol{\beta}, D_b, h_b)\big\} + \lambda_b \Vert \boldsymbol{\beta}_{-0} \Vert_1 \bigg\},$$ 
    where $\lambda_b$ is tuned by 5-fold cross-validation.
    
    \STATE 2: Compute $\nabla \widehat{L}(\widehat{\boldsymbol{\beta}}_b^{\mathrm{on}}, D_b, h_b)$ and $\nabla^2 \widehat{L}(\widehat{\boldsymbol{\beta}}_b^{\mathrm{on}}, D_b, h_b)$.   
    
    \STATE 3: Update the summary statistics $\boldsymbol{S}_{b,1}$, $\boldsymbol{S}_{b,2}$, $\widehat{\boldsymbol{H}}_b$, and $\widehat{{\boldsymbol{\Sigma}}}_{b}$ in an online manner as (\ref{summary statistics update case}) .
    
    \STATE 4: Estimate the precision matrix $\widehat{\boldsymbol{H}}_b^{-1}$ by $\widehat{\boldsymbol{\Theta}}_{b}$, where the $j$th-column of $\widehat{\boldsymbol{\Theta}}_{b}$ is defined by 
    $\widehat{\boldsymbol{\Theta}}_{bj} = {\text{arg min}}_{\boldsymbol{\Theta}_{bj} \in \mathbb{R}^{p+1}}\Vert \boldsymbol{\Theta}_{bj} \Vert _1, \:\text{s.t.} \Vert \widehat{\boldsymbol{H}}_b \boldsymbol{\Theta}_{bj} - \boldsymbol{e}_j \Vert _{\infty} \leqslant \delta_{b}$, and $\delta_b$ is tuned by 5-fold cross-validation.
    
    \STATE 5: Calculate the debiased estimate: $$\widehat{\boldsymbol{\beta}}_{b,j}^{\mathrm{on},\mathrm{de}}  = \widehat{\boldsymbol{\beta}}_{b,j}^{\mathrm{on}} - \frac{1}{N_b}(\boldsymbol{\widehat{\Theta}}_{bj})^{\top} \boldsymbol{S}_{b,1} + \frac{1}{N_b}(\boldsymbol{\widehat{\Theta}}_{bj})^{\top} \boldsymbol{S}_{b,2} - (\boldsymbol{\widehat{\Theta}}_{bj})^{\top} \widehat{\boldsymbol{H}}_{b} \widehat{\boldsymbol{\beta}}_b^\mathrm{on}.$$
    
    \STATE 6: Estimate the variance of the debiased estimate $\widehat{\boldsymbol{\beta}}_{b,j}^{\mathrm{on},\mathrm{de}}$ by $\widehat{{\sigma}}_{bj}^2 = \widehat{\boldsymbol{\Theta}}_{bj}^\top  \widehat{\boldsymbol{\Sigma}}_{b} \widehat{\boldsymbol{\Theta}}_{bj}$.
    
    \STATE 7: Construct 95\% confidence interval: $$\left(\widehat{\boldsymbol{\beta}}_{b,j}^{\mathrm{on},\mathrm{de}} - 1.96 \cdot\widehat{{\sigma}}_{bj} / \sqrt{N_b}, \widehat{\boldsymbol{\beta}}_{b,j}^{\mathrm{on},\mathrm{de}}  + 1.96 \cdot\widehat{{\sigma}}_{bj} / \sqrt{N_b}\right).$$
    
    \STATE 8: Output the debiased Lasso estimate $\widehat{\boldsymbol{\beta}}_{b,j}^{\mathrm{on},\mathrm{de}}$ and its 95\% confidence interval for the $b$-th batch.
\ENDFOR
\end{algorithmic}
\end{algorithm}

\section{Theoretical properties\label{theory}}
\subsection{Properties of the offline estimator \label{offline theory}} 

In this subsection, we aim to provide theoretical support for our offline method. Specifically, we first show the finite-sample error bound of the Lasso estimate under both \(l_1\) and \(l_2\) norms, i.e., \(\|\widehat{\boldsymbol{\beta}}^{\mathrm{off}}-\boldsymbol{\beta}^*\|_1\) and \(\|\widehat{\boldsymbol{\beta}}^{\mathrm{off}}-\boldsymbol{\beta}^*\|_2\), where \(\widehat{\boldsymbol{\beta}}^{\mathrm{off}}\) is defined in (\ref{offline smooth lasso estimator}). Second, we demonstrate the finite-sample bound of the estimate of the inverse Hessian matrix, \(\| \widehat{\boldsymbol{\Theta}} - \boldsymbol{\Theta}^* \|_{L_1}\), where \(\widehat{\boldsymbol{\Theta}}\) is defined in (\ref{offline information estimation}) and \(\boldsymbol{\Theta}^*\) is defined as $\boldsymbol{\Theta}^* =  [E\{\nabla^2 l_h(Y\tilde{\boldsymbol{X}}^{\top}\boldsymbol{\beta}^*)\}]^{-1}$. Third, by combining the first two results, we derive our final result of interest, the asymptotic normality of the debiased Lasso estimate, \(\widehat{\boldsymbol{\beta}}_j^{\mathrm{off},\mathrm{de}}\), and therefore validate the confidence interval defined in (\ref{offline interval}).

Recall that $\tilde{\boldsymbol{X}} = (1, X_1, \ldots, X_p)^\top$. Let $\tilde{\boldsymbol{X}}_{-1}$ be a $p$-dimensional vector with $X_1$ removed from $\tilde{\boldsymbol{X}}$.
Similar notations are used for $\boldsymbol{\beta}$. Let $f$ and $g$ be the probability density function of $\tilde{\boldsymbol{X}}$ when $Y = 1$ and $Y = -1$, respectively. Let $f(x|\tilde{\boldsymbol{X}}_{-1})$ be the conditional density function of $X_1$ given $\tilde{\boldsymbol{X}}_{-1}$ when \(Y=1\) and $f_{-1}(\tilde{\boldsymbol{X}}_{-1})$ be the joint density of $\tilde{\boldsymbol{X}}_{-1}$ when \(Y=1\). Similar notations are used for $g(\cdot)$ where \(Y=-1\). We consider the assumptions below:
\begin{assumption}\label{assump:A1}
The kernel \(K(t) \in [0,\infty)\) is symmetric and satisfies \(\int_{-\infty}^{\infty} K(t)dt = 1\) as well as \(\int_{-\infty}^{\infty} t K(t)dt = 0\). Moreover, it fulfills the following conditions:
(i) \(\kappa_u' = \sup_{t\in\mathbb{R}} |K^{(1)}(t)| < \infty\) and \(\kappa_u'' = \sup_{t\in\mathbb{R}} |K^{(2)}(t)| < \infty\);
(ii) \(\kappa_u = \sup_{t\in\mathbb{R}} K(t) < \infty\), and \(\kappa_l = \min_{|t|\le 1} K(t) > 0\);
(iii) \(\kappa_1, \kappa_2 < \infty\), where \(\kappa_a = \int_{-\infty}^{\infty} |t|^a K(t)dt\) for any \(a>0\).

\end{assumption}
\begin{assumption}\label{assump:A2}
   The density function $f(\boldsymbol{x})$ is continuous, bounded, and possesses finite second moments. Assume that $\sup_{x \in \mathbb{R}} \max \{|f(x|\tilde{\boldsymbol{X}}_{-1})|, |f'(x|\tilde{\boldsymbol{X}}_{-1})|, |x^2 f(x|\tilde{\boldsymbol{X}}_{-1})|, |x f'(x|\tilde{\boldsymbol{X}}_{-1})|$, \\
   $ |x^2 f'(x|\tilde{\boldsymbol{X}}_{-1})| \} \leq C_f$ for some constant $C_f > 0$. Also assume $\int_{\mathbb{R}} |x| f(x|\tilde{\boldsymbol{X}}_{-1}) dx < \infty$. Similar assumptions are made for $g(\cdot)$.
\end{assumption}

\begin{assumption}\label{assump:A3}
Let $\boldsymbol{\boldsymbol{\Sigma}}=E[\tilde{\boldsymbol{X}} \tilde{\boldsymbol{X}}^\top]$. We assume that \(\boldsymbol{\boldsymbol{\Sigma}}\) satisfies the following conditions: (i) there exists a constant $v_0 > 0$ such that, for any vector $\boldsymbol{u} \in \mathbb{R}^{p + 1}$ and $t > 0$, $Pr(|\boldsymbol{u}^T \boldsymbol{\omega}| \geqslant \nu_0 \Vert \boldsymbol{u} \Vert_2 t) \leq 2 \, \mathrm{exp}(-t^2 / 2)$, where $\boldsymbol{\omega} = \boldsymbol{\boldsymbol{\Sigma}}^{-1/2}\tilde{\boldsymbol{X}}$; (ii) $0 < c_1 \leq \gamma_{p+1} = \gamma_{p+1}(\boldsymbol{\boldsymbol{\Sigma}}) \leq \gamma_1 = \gamma_1(\boldsymbol{\boldsymbol{\Sigma}}) \leq c_2 < \infty$ for constants $c_1, c_2$;
(iii) there exists a constant M such that $\|\boldsymbol{\boldsymbol{\Sigma}}^{-1}\|_{L_1} \leq M$.

\end{assumption}

\begin{assumption} \label{assump: true parameter bound}
    $|\boldsymbol{\beta}_1^*| \geq c$, $\| \boldsymbol{\beta}^* \|_2 \leq C$ for constants $c, C > 0$.
\end{assumption}

\begin{assumption} \label{assump: hessian bound}
    Define $D = E[\delta(1 - Y\tilde{\boldsymbol{X}}^\top \boldsymbol{\beta}^*)\tilde{\boldsymbol{X}}\tilde{\boldsymbol{X}}^\top]$, where $\delta(\cdot)$ is the Dirac delta function, assume $D$ satisfies $0 < a_1 \leq \gamma_{p+1}(D) \leq \gamma_1(D) \leq a_2 < \infty$ for constants $a_1, a_2$. 
\end{assumption}

These assumptions are commonly utilized in the related literature \cite[]{cai2016assump, tan2022high, wang2025convolution, xie2025statistical, chen2025efficient}. Assumption \ref{assump:A1} posits conditions for the kernel function \(K(t)\) utilized to smooth the hinge loss. This assumption is satisfied by the Gaussian kernel used in our paper, and also by some other kernels frequently utilized, such as the logistic kernel. 
Assumption \ref{assump:A2} imposes some boundedness conditions on the density functions $f(\cdot)$ and $g(\cdot)$, which are satisfied by commonly used distributions such as the Gaussian distribution and the uniform distribution.
Without loss of generality, Assumption \ref{assump:A3} specifies some regularity conditions for the predictors \(\boldsymbol{X}\). Assumption \ref{assump:A3}(i) supposes that \(\boldsymbol{X}\) is sub-Gaussian. Assumption \ref{assump:A3}(ii) assumes that the eigenvalue of the matrix \(\boldsymbol{\boldsymbol{\Sigma}}=E[\tilde{\boldsymbol{X}} \tilde{\boldsymbol{X}}^\top]\) is bounded away from zero and infinity such that it is invertible. Assumption \ref{assump:A3}(iii) states that \(\|\boldsymbol{\boldsymbol{\Sigma}}^{-1}\|_{L_1}\) is bounded, which is satisfied by many covariance structures of \(\boldsymbol{\boldsymbol{\Sigma}}\), e.g., the autoregressive correlation structure. 
Assumption \ref{assump: true parameter bound} supposes that \(X_1\) is an important variable of the decision boundary \(\tilde{\boldsymbol{X}}^T\boldsymbol{\beta}^*\), i.e., \(|\boldsymbol{\beta}_1^*|\geq 0\), and that $\|\boldsymbol{\beta}^*\|_2$ is bounded. Assumption \ref{assump: hessian bound} requires that the eigenvalues of the population Hessian matrix  $D$ are bounded away from zero and infinity. 

\begin{theorem} 
\label{theorem: offline beta consistency}
    Under Assumptions \ref{assump:A1}-\ref{assump: hessian bound}, by choosing $\lambda \asymp \sqrt{\log(p)/n}$ and $h \asymp (s \log(p)/n)^{1/4}$, there exist constants $C_1$ and $C_2$ (depending on $\gamma_{p+1}, \gamma_1, C_f, c, \kappa_2, \kappa_l, a_1, a_2$) such that with probability greater than $1 -( 4/p)$, 
    $$
    \|\widehat{\boldsymbol{\beta}}^{\mathrm{off}} - \boldsymbol{\beta}^*\|_2 \le C_2 s^{1/2} \lambda , \quad \|\widehat{\boldsymbol{\beta}}^{\mathrm{off}} - \boldsymbol{\beta}^*\|_1 \le C_1 s \lambda.
    $$ 
\end{theorem}

Theorem \ref{theorem: offline beta consistency} provides finite sample upper bounds for the offline Lasso estimate $\widehat{\boldsymbol{\beta}}^{\mathrm{off}}$ under both the $l_1$ and $l_2$ norms. It also indicates that, when we choose $\lambda \asymp \sqrt{ \mathrm{log}(p)/n}$ and $h \asymp (s \log(p)/n)^{1/4}$, our Lasso estimate can achieve a convergence rate of $O_p(s\sqrt{\log (p)/n})$ under the $l_1$-norm and \(O_p(\sqrt{s \log (p)/n})\) under the $l_2$-norm. These convergence rates are the same as the Lasso SVM estimate without smoothing \cite[]{peng2016error}. Therefore, with a proper choice of the smoothing parameter \(h\), the kernel smoothing does not deteriorate the convergence rates, which is an appealing property of the kernel smoothing.

To establish the finite sample bound of the estimate of the inverse Hessian matrix \(\boldsymbol{\Theta}^*\), we further require the following assumption.

\begin{assumption}
    \label{assumption: offline inverse matrix}  
    Assume there exist positive $R_{1, h, p}, R_{2, h, p}$ such that $\boldsymbol{\Theta}^*$ satisfies: (i) $\| \boldsymbol{\Theta}^* \|_{L_1} \leq R_{1, h, p}$ (ii) $\| \boldsymbol{\Theta}^{*} \|_{L_q} \leq R_{2, h, p}$ for any $q \in [0, 1)$.
\end{assumption}

Analogous assumptions have been widely adopted in the literature of high-dimensional inverse Hessian matrix estimation \cite[]{clime, fan2016sparse_assumption, lian2018divide}. 
Assumption \ref{assumption: offline inverse matrix} requires that the $L_{q}$-norm for \(q\in[0,1]\) of the precision matrix $\boldsymbol{\Theta}^{*}$ to be bounded from \(R_{1,h,p}\) and \(R_{2,h,p}\). Since \(R_{1,h,p}\) and \(R_{2,h,p}\) can depend on the feature dimension \(p\) and are therefore allowed to diverge as \(p\) grows, this assumption is relatively mild.

By combining Assumption \ref{assumption: offline inverse matrix} and Theorem \ref{theorem: offline beta consistency}, we can derive the finite sample error bound of $\|\widehat{\boldsymbol{\Theta}}-\boldsymbol{\Theta}^*\|_{L_1}$ as follows.

\begin{theorem}
    \label{theorem: offline theta consistency}
    Suppose the conditions in Theorem \ref{theorem: offline beta consistency} and Assumption \ref{assumption: offline inverse matrix} hold. Let
    \begin{align*}
        \delta = \sqrt{\frac{\log p}{n h}} + \frac{\log p}{n h} + \frac{C_{1} s \log p }{h \sqrt{n}} + \frac{C_{1} s (\log p)^2 }{h^2 n^{3/2}} + \frac{C_{1}^2 s^2 (\log p)^3}{h^3 n} + \frac{ C_{1} s (\log p)^{3/2}}{n h^{3/2}}, 
    \end{align*}
    then, there exists a constant $C_3 > 0$ depending on $\kappa_u, \kappa_u^{'}, \kappa_u^{''}, C_f, c$, such that for any $q \in [0, 1)$, with probability greater than $1 - (4/p)$,
    $$
    \| \widehat{\boldsymbol{\Theta}} - \boldsymbol{\Theta}^* \|_{L_1} \leq 2C_3 R_{2, h, p}(\delta \| \boldsymbol{\Theta}^* \|_{L_1})^{1-q} \asymp R_{2, h, p} (R_{1, h, p}\delta)^{1-q}.
    $$
\end{theorem}
     
Theorem \ref{theorem: offline theta consistency} establishes a finite sample upper bound for the estimation error $\| \widehat{\boldsymbol{\Theta}} - \boldsymbol{\Theta}^* \|_{L_1}$. Since $h \asymp (s \log (p)/n)^{1/4}$ as in Theorem \ref{theorem: offline beta consistency}, by letting $q = 0$, we can obtain a convergence rate for the estimate of the inverse Hessian matrix, i.e., $ \| \widehat{\boldsymbol{\Theta}} - \boldsymbol{\Theta}^* \|_{L_1} = O_p(R_{2, h, p} R_{1, h, p} s^{5/4} (\log p)^{9/4} n^{-1/4})$. Consequently, the estimate $\widehat{\boldsymbol{\Theta}}$ is consistent to $\boldsymbol{\Theta}^*$ provided that $R_{1,h,p}^4R_{2,h,p}^4  = o[n/\{s^5(\log p)^9\}]$, where \(R_{1,h,p},R_{2,h,p}\) are upper bound of $\| \boldsymbol{\Theta}^{*} \|_{L_1}$ and $\| \boldsymbol{\Theta}^{*} \|_{L_0}$, respectively. This is mild since \(R_{1,h,p}R_{2,h,p}\) can diverge as \(n\to\infty\). Our requirement for \(R_{1,h,p}R_{2,h,p}\) is stronger than those methods that do not utilize kernel smoothing, such as \cite{debias_lasso_vande, tiger_s_choice}, which require that 
\(R_{1,h,p}R_{2,h,p}=o[\{n/\log p\}^{1/2}]\). This difference arises from the fact that we should further handle the kernel smoothing procedure, which introduces smoothing bias to the parameter estimation. 

By combining the results in Theorems \ref{theorem: offline beta consistency} and \ref{theorem: offline theta consistency}, we can propose a Bahadur representation of our debiased Lasso estimate \(\widehat{\boldsymbol{\beta}}^{\mathrm{off},\mathrm{de}}_j\) and therefore establish the asymptotic normality of the debiased Lasso estimate:
\begin{theorem}
    \label{theorem: offline asymptotic}
    Assume conditions in Theorems \ref{theorem: offline beta consistency} and \ref{theorem: offline theta consistency} hold, under scaling condition $R_{1, h, p}R_{2, h, p}C_{1}^3s^3(\log p)^{7/2} /(n h^3) = o(1)$, we obtain a Bahadur representation of $\widehat{\boldsymbol{\beta}}^{\mathrm{off},\mathrm{de}}_j$:
    \begin{align*}
         \sqrt{n}(\widehat{\boldsymbol{\beta}}^{\mathrm{off},\mathrm{de}}_j - \boldsymbol{\beta}_j^*) = Z_j^{\mathrm{off}} + W_j^{\mathrm{off}},\ j=1,...,p+1,
    \end{align*}
    where $Z_j^{\mathrm{off}} = -\frac{1}{\sqrt{n}} (\boldsymbol{\Theta}_{j}^*)^T \sum_{i = 1}^{n} \{-\mathcal{K}_h(1-Y_i\tilde{\boldsymbol{X}}_i^T {\boldsymbol{\beta}}^*)Y_i\tilde{\boldsymbol{X}}_i \}$ is asymptotically normal and $ |W_j^\mathrm{off}|=o_p(1)$ is asymptotically negligible. Here, \(\boldsymbol{\Theta}_{j}^*\) is the \(j\)-th column of $\boldsymbol{\Theta}^*$.
\end{theorem}

Theorem \ref{theorem: offline asymptotic} states that \( \sqrt{n}(\widehat{\boldsymbol{\beta}}^{\mathrm{off},\mathrm{de}}_j - \boldsymbol{\beta}_j^*)\) can be decomposed into a sum of \(Z_j^{\mathrm{off}}\) and \( W_j^{\mathrm{off}}\). The \(Z_j^{\mathrm{off}}\) converges in distribution to a normal distribution with mean \(E\{Z_j^{\mathrm{off}}\}=0\) and variance $\sigma_j^2 = (\boldsymbol{\Theta}_{j}^*)^\top \boldsymbol{\Sigma}^* \boldsymbol{\Theta}_{j}^*$ ,where $\boldsymbol{\Sigma}^* = E[\tilde{\boldsymbol{X}} \tilde{\boldsymbol{X}}^\top \{\mathcal{K}_h(1 - Y \tilde{\boldsymbol{X}}^\top \boldsymbol{\beta}^*)\}^2 ]$. Meanwhile, the remainder term $ W_j^{\mathrm{off}}$ converges to $0$ in probability. Therefore, by utilizing Slutsky's theorem, we obtain that 

$$
\sqrt{n}(\widehat{\boldsymbol{\beta}}^{\mathrm{off},\mathrm{de}}_j - \boldsymbol{\beta}_j^*) \xrightarrow{d} \mathcal{N}(0, \sigma_j^2).
$$

Consequently, to derive a valid confidence interval based on the debiased Lasso estimate \(\widehat{\boldsymbol{\beta}}^{\mathrm{off},\mathrm{de}}_j\), we only need a consistent estimate of the variance \(\sigma_j^2=(\boldsymbol{\Theta}_{j}^*)^\top \boldsymbol{\Sigma}^* \boldsymbol{\Theta}_{j}^*\). We propose \(\widehat{{\sigma}}_{j}^2 = (\widehat{\boldsymbol{\Theta}}_{j})^\top \widehat{\boldsymbol{\boldsymbol{\Sigma}}} \widehat{\boldsymbol{\Theta}}_{j}\) as defined in (\ref{sigma: offline sigma}), and construct a 95\% Wald-type confidence interval for $\widehat{\boldsymbol{\beta}}^{\mathrm{off},\mathrm{de}}_j$ by $\big(\widehat{\boldsymbol{\beta}}^{\mathrm{off},\mathrm{de}}_j - 1.96 \cdot\widehat{{\sigma}}_{j} / \sqrt{n}, \widehat{\boldsymbol{\beta}}^{\mathrm{off},\mathrm{de}}_j  + 1.96 \cdot\widehat{{\sigma}}_{j} / \sqrt{n}\big)$. 
The \(\widehat{{\sigma}}_{j}^2\) is consistent to \(\sigma_j^2\) since \(\| \widehat{\boldsymbol{\Theta}} - \boldsymbol{\Theta}^{*} \|_{\infty}\) and \(\| \widehat{\boldsymbol{\Sigma}} - \boldsymbol{\Sigma}^{*} \|_{\infty}\) converges in probability to 0 as follow:

\begin{corollary}
\label{corollary: Sigma and Theta consistency}
    Assume conditions in Theorem \ref{theorem: offline asymptotic} hold, we have
    $$
    \| \widehat{\boldsymbol{\Theta}} - \boldsymbol{\Theta}^* \|_{\infty} \lesssim R_{1, h, p} ^2\delta \quad \text{and} \quad \| \widehat{\boldsymbol{\Sigma}} - \boldsymbol{\Sigma}^* \|_{\infty} \lesssim \delta + \sqrt{\frac{2 \log p}{n}} + \frac{\log p}{n}
    $$
    hold with probability $1 - (4/p)$. 
\end{corollary}

By replacing \(\delta\) in Corollary \ref{corollary: Sigma and Theta consistency} with its definition in Theorem \ref{theorem: offline theta consistency} and under the scaling condition in Theorem \ref{theorem: offline asymptotic}, we obtain that $\| \widehat{\boldsymbol{\Theta}} - \boldsymbol{\Theta}^* \|_{\infty}=o_p(1)$ and $ \| \widehat{\boldsymbol{\Sigma}} - \boldsymbol{\Sigma}^* \|_{\infty}=o_p(1)$. 

\subsection{Properties of the online estimator \label{online theory}}
In this subsection, we develop the theoretical guarantees for our online method. We first bound the finite-sample estimation error of the online Lasso estimate for the $b$-th batch under both \(l_1\)- and \(l_2\)-norm, i.e., \(\|\widehat{\boldsymbol{\beta}}^{\mathrm{on}}_b-\boldsymbol{\beta}^*\|_1\) and \(\|\widehat{\boldsymbol{\beta}}^{\mathrm{on}}_b-\boldsymbol{\beta}^*\|_2\), where \(\widehat{\boldsymbol{\beta}}^{\mathrm{on}}_b\) is defined in (\ref{online lasso bias estimator}). Next, we show the finite sample error bound for the estimate of the inverse Hessian matrix, \(\| \widehat{\boldsymbol{\Theta}}_b - \boldsymbol{\Theta}^* \|_{L_1}\). Third, by leveraging these two bounds, we establish the asymptotic normality for the debiased online Lasso estimate, \(\widehat{\boldsymbol{\beta}}_{b,j}^{\mathrm{on},\mathrm{de}}\), and therefore validate the confidence interval defined in (\ref{online interval}).

\begin{theorem} \label{theorem:online bound}
    Under Assumptions \ref{assump:A1}-\ref{assump: hessian bound}, let $\lambda_b \asymp \sqrt{\log(p)/N_b}$ and $h_b \asymp \{s \log(p)/N_b\}^{1/4}$  for $b = 1, 2, \ldots, B$. For $b = 2, 3, \ldots, B$, define $C_{b,1} = \big(4 + 2C_f \kappa_2/c + 2 a_2 C_{b-1, 2}\big)C_{b,2}$ and $C_{b,2}=\big(3 + 2C_f \kappa_2/c + 2 a_2 C_{b-1, 2}\big)\gamma_1 \gamma_{p+1}^{-1}a_2 / (\kappa_l a_1^2)$ with $C_{1,1} = \big(4 + 2C_f \kappa_2/c\big)C_{1,2}$ and $C_{1, 2} = \big(3 + 2C_f \kappa_2/c \big)\gamma_1 \gamma_{p+1}^{-1}a_2 / (\kappa_l a_1^2)$. Under the scaling condition $C_{b-1, 1}^2 s^2 (\log p)^3 \{ 1 + \log (N_{b-1}/n_1) \} = o(N_{b-1}h_{b-1}^3)$, with probability greater than $1-(4/p)$,
    $$
    \|\widehat{\boldsymbol{\beta}}_b^{\mathrm{on}} - \boldsymbol{\beta}^*\|_2 \le C_{b, 2} s^{1/2} \lambda_b, \quad \|\widehat{\boldsymbol{\beta}}_b^{\mathrm{on}} - \boldsymbol{\beta}^*\|_1 \le C_{b, 1} s \lambda_b.
    $$
\end{theorem}

Theorem \ref{theorem:online bound} establishes finite sample error bounds for the online Lasso estimate $\widehat{\boldsymbol{\beta}}^{\mathrm{on}}_b$ under the $l_1$- and $l_2$-norms. The terms \(C_{b, 1},C_{b, 2}\) grow exponentially with the time points \(b\), e.g., $C_{b, 1} \asymp (1 + A_1)^{2(b-1)}$ and $C_{b, 2} \asymp (1 + A_2)^{b-1}$ for some constants \(A_1,A_2>0\). Therefore, the upper bound in Theorem \ref{theorem:online bound} deteriorates as \(b\) increases. This arises from the propagation of approximation errors incurred by relying on summary statistics from previous batches rather than individual observations in constructing our online Lasso estimate. Such error accumulation is a well-documented phenomenon for online updating methods \cite[]{luo2023online, xie2025statistical}.

Though the approximation error is accumulated as the time point \(b\) increases, the asymptotic behavior of the online Lasso estimate $\widehat{\boldsymbol{\beta}}^{\mathrm{on}}_b$ is comparable with the offline Lasso estimate $\widehat{\boldsymbol{\beta}}^{\mathrm{off}}_b$ under certain conditions for \(b\). Specifically, when $b$ is fixed, the terms $C_{b, 1}$ and $C_{b, 2}$ are bounded constants. Consequently, our online Lasso estimate achieves a convergence rate of $O_p(\sqrt{s \log (p)/N_b})$ under the $l_2$-norm and $O_p(s\sqrt{ \log (p)/N_b})$ under the $l_1$-norm, which are the same as those of the offline case in Theorem \ref{theorem: offline beta consistency}. When \(b\) grows with sample sizes \(N_b\) such that \(b=o(\log N_b)\), our online Lasso estimate attains a convergence rate of $O_p(\sqrt{s \log (p)/N_b^{1-\epsilon}})$ under the $l_2$-norm and $O_p(s\sqrt{ \log (p)/N_b^{1-\epsilon}})$ under the $l_1$-norm for any \(\epsilon>0\). This indicates that the convergence rates of the online Lasso estimate are only slightly slower than that of the offline case in Theorem \ref{theorem: offline beta consistency}.

By combining Assumption \ref{assumption: offline inverse matrix} and Theorem \ref{theorem:online bound}, we can derive the finite sample error bound of $\|\widehat{\boldsymbol{\Theta}}_b-\boldsymbol{\Theta}^{*}\|_{L_1}$ as follows.

\begin{theorem}
    \label{theorem:inverse and expectation}
    Suppose that the conditions in Theorem \ref{theorem:online bound} and Assumption \ref{assumption: offline inverse matrix} hold for $b = 1, 2, \ldots, B$. Let
    \begin{align*}
        \delta_b &= \sqrt{\frac{\log p}{N_b h_b}} + \frac{\log p}{N_b h_b} + \frac{s \log (p) C_{b, 1}}{h_b \sqrt{N_b}} + \frac{sb (\log p)^2 C_{b, 1}}{h_b^2 N_b \sqrt{n_1}} \\
        &\quad + \frac{C_{b, 1}^2 s^2 (\log p)^3 \{ 1 + \log(N_b/n_1) \}}{h_b^3 N_b} + \frac{s (\log p)^{3/2} C_{b, 1} \sqrt{b + b \log(N_b/n_1)}}{N_b(h_b)^{3/2}}, 
    \end{align*}
    then, there exists $C_4 > 0$ depending only on $\kappa_u, \kappa_u^{'}, \kappa_u^{''}, C_f, c$, such that for any $q \in [0, 1)$, with probability greater than $1 - (4/p)$,
    $$
   \| \widehat{\boldsymbol{\Theta}}_b - \boldsymbol{\Theta}^* \|_{L_1} \leq 2C_4 R_{2, h, p}(\delta_b \| \boldsymbol{\Theta}^* \|_{L_1})^{1-q} \asymp R_{2, h, p} (R_{1, h, p}\delta_b)^{1-q}.
    $$
\end{theorem}

Theorem \ref{theorem:inverse and expectation} establishes a finite sample upper bound for the online estimate $\widehat{\boldsymbol{\Theta}}_b$. Since $h_b \asymp (s \log (p)/N_b)^{1/4}$ as in Theorem \ref{theorem:online bound}, by setting $q = 0$, we can derive the convergence rate of \(\widehat{\boldsymbol{\Theta}}_b\) as $\| \widehat{\boldsymbol{\Theta}}_b - \boldsymbol{\Theta}^{*} \|_{L_1} = O_p(R_{2, h, p} R_{1, h, p} C_{b, 1}^2 s^{5/4} (\log p)^{9/4} N_b^{-1/4} \{ 1 + \log(N_b/n_1) \})$. Consequently, consistency is achieved provided that $R_{1,h,p}^4 R_{2,h,p}^4 = o(N_b / [C_{b, 1}^8 s^5 (\log p)^9 \{ 1 + \log(N_b/n_1) \}])$. For simplicity, we assume that the sample sizes of different batches are equal, i.e., \(n_1=n_2= \ldots =n_b\). Then, when \(\log(N_b/n_1)\asymp1\), which indicates that the time point \(b\) is a constant that do not diverge to \(\infty\), the consistency requirement becomes $R_{1,h,p}^4 R_{2,h,p}^4 = o(N_b / [s^5 (\log p)^9])$, which is the same as the requirement of the offline estimate \(\widehat{\boldsymbol{\Theta}}\) to achieve consistency discussed under Theorem \ref{theorem: offline theta consistency}. When the time points \(b\to\infty\), the requirement of consistency for the online estimate \(\widehat{\boldsymbol{\Theta}}_b\) is stronger than that of the offline estimate in Theorem \ref{theorem: offline theta consistency}. In particular, when \(b=o(\log N_b)\), the requirement for consistency becomes $R_{1,h,p}^4 R_{2,h,p}^4 = o(N_b^{1-\epsilon} / [s^5 (\log p)^9])$ for any \(\epsilon>0\). This is only slightly stronger than the requirement for offline estimate in Theorem \ref{theorem: offline theta consistency}.

By combining the results in Theorems \ref{theorem:online bound} and \ref{theorem:inverse and expectation}, we can establish a Bahadur representation of the debiased Lasso estimate $\widehat{\boldsymbol{\beta}}_{b,j}^{\mathrm{on},\mathrm{de}}$ and establish its asymptotic normality as follows:

\begin{theorem}
    \label{theorem:debias theory}
    Assume conditions in Theorem \ref{theorem:online bound} and \ref{theorem:inverse and expectation} hold, under scaling condition $R_{2, h, p}R_{1, h, p}C_{b,1}^3s^3(\log p)^{7/2} \{ 1 + \log(N_b/n_1) \}/(N_b h_b^3) = o(1)$, then for $j = 1,2, \ldots,p+1$, we obtain a Bahadur representation of $\widehat{\boldsymbol{\beta}}_{b,j}^{\mathrm{on},\mathrm{de}}$:
    
     \begin{align*}
    \sqrt{N_b}(\widehat{\boldsymbol{\beta}}_{b,j}^{\mathrm{on},\mathrm{de}} - \boldsymbol{\beta}_j^*) = Z_j^{\mathrm{on}} + W_j^{\mathrm{on}},
    \end{align*}
    where $Z_j^{\mathrm{on}} = -\frac{1}{\sqrt{N_b}} (\boldsymbol{\Theta}_{j}^*)^T \sum_{t=1}^b \sum_{i = 1}^{n_t} \{-\mathcal{K}_{h_t}(1-Y_{ti}\tilde{\boldsymbol{X}}_{ti}^T \boldsymbol{\beta}^*)Y_{ti}\tilde{\boldsymbol{X}}_{ti} \}$ is asymptotically normal and $|W_j^{\mathrm{on}}| = o_p(1)$ is asymptotically negligible.
\end{theorem}

Theorem \ref{theorem:debias theory} decomposes \( \sqrt{N_b}(\widehat{\boldsymbol{\beta}}_{b,j}^{\mathrm{on},\mathrm{de}} - \boldsymbol{\beta}_j^*)\) into \(Z_j^{\mathrm{on}}\) and \( W_j^{\mathrm{on}}\). The \(Z_j^{\mathrm{on}}\) is asymptotically normal with mean \(E\{Z_j^{\mathrm{on}}\}=0\) and variance $\sigma_{j}^2 = (\boldsymbol{\Theta}_{j}^*)^\top \boldsymbol{\Sigma}^* \boldsymbol{\Theta}_{j}^*$
, while the remainder $ W_j^{\mathrm{on}}$ converges to $0$ in probability. Then we can obtain that 
$$
\sqrt{N_b}(\widehat{\boldsymbol{\beta}}_{b,j}^{\mathrm{on},\mathrm{de}} - \boldsymbol{\beta}_j^*) \xrightarrow{d} \mathcal{N}(0, \sigma_{j}^2).
$$

To conduct statistical inference, we propose \(\widehat{{\sigma}}_{bj}^2 = (\widehat{\boldsymbol{\Theta}}_{bj})^\top \widehat{\boldsymbol{\boldsymbol{\Sigma}}}_b \widehat{\boldsymbol{\Theta}}_{bj}\) as defined in (\ref{sigma: online sigma}) as an estimate of $\sigma_{j}^2$, and construct a 95\% Wald-type confidence interval for $\widehat{\boldsymbol{\beta}}_{b,j}^{\mathrm{on},\mathrm{de}}$ by $\big(\widehat{\boldsymbol{\beta}}_{b,j}^{\mathrm{on},\mathrm{de}}- 1.96 \cdot\widehat{{\sigma}}_{bj} / \sqrt{N_b}, \widehat{\boldsymbol{\beta}}_{b,j}^{\mathrm{on},\mathrm{de}}  + 1.96 \cdot\widehat{{\sigma}}_{bj} / \sqrt{N_b}\big)$. The consistency of $\widehat{{\sigma}}_{bj}^2$ to ${{\sigma}}_{j}^2$ is guaranteed by the convergence of its components, $\| \widehat{\boldsymbol{\Theta}}_b - \boldsymbol{\Theta}^* \|_{\infty}$ and $\| \widehat{\boldsymbol{\Sigma}}_b - \boldsymbol{\Sigma}^* \|_{\infty}$, as detailed below:
\begin{corollary}
\label{corollary: online Sigma and Theta consistency}
    Assume conditions in Theorem \ref{theorem:debias theory} hold, we have
    $$
    \| \widehat{\boldsymbol{\Theta}}_b - \boldsymbol{\Theta}^* \|_{\infty} \lesssim R_{1, h, p} ^2\delta_b \quad \text{and} \quad \| \widehat{\boldsymbol{\Sigma}}_b - \boldsymbol{\Sigma}^* \|_{\infty} \lesssim \delta_b + \sqrt{\frac{2 \log p}{N_b}} + \frac{\log p}{N_b}
    $$
    hold with probability $1 - (4/p)$.
\end{corollary}

By substituting the definition of $\delta_b$ from Theorem \ref{theorem:inverse and expectation} into Corollary \ref{corollary: online Sigma and Theta consistency}, and applying the scaling conditions in Theorem \ref{theorem:debias theory}, we conclude that $\| \widehat{\boldsymbol{\Theta}}_b - \boldsymbol{\Theta}^* \|_{\infty} = o_p(1)$ and $\| \widehat{\boldsymbol{\Sigma}}_b - \boldsymbol{\Sigma}^* \|_{\infty} = o_p(1)$.

\section{Simulation studies \label{simulation}}
In this section, we explore the finite sample performance of our methods through simulation. We follow a traditional data generation method in \cite[]{lian2018divide, peng2016error, chen2025efficient}. Specifically, assume the response \(Y\) are equiprobable, $P(Y=1) = P(Y=-1) = 0.5$, and the features \(\boldsymbol{X}\) are distributed as $\boldsymbol{X} \mid (Y = 1) \sim \mathcal{N}_p(\boldsymbol{\mu}, \boldsymbol{\boldsymbol{\Sigma}})$ and $\boldsymbol{X} \mid (Y = -1) \sim \mathcal{N}_p(\boldsymbol{\nu}, \boldsymbol{\boldsymbol{\Sigma}})$. Here, we consider two choices of \(\boldsymbol{\mu},\boldsymbol{\nu}\) and three choices of \(\boldsymbol{\Sigma}\), yielding six scenarios as follow: 

\begin{case} \label{case1}
We set $\boldsymbol{\mu} = (0.3, 0.3, 0.3, 0.3, 0.3, 0, \ldots, 0)^\top \in \mathbb{R}^p$, $\boldsymbol{\nu} = -\boldsymbol{\mu}$. We examine three different covariance structures:
(1) type I: $\boldsymbol{\boldsymbol{\Sigma}} = (\boldsymbol{\Sigma}_{ij})_{1 \leq i, j \leq p} \in \mathbb{R}^{p \times p}$ with $\boldsymbol{\Sigma}_{ii} = 1$ for all $i$, $\boldsymbol{\Sigma}_{ij} = 0.1$ if $i \neq j$ and $i,j \leq 5$, and $\boldsymbol{\Sigma}_{ij} = 0$ otherwise; (2) type II: A block-diagonal covariance matrix composed of two submatrices, $\boldsymbol{\boldsymbol{\Sigma}}_{s \times s}$ and $\boldsymbol{\boldsymbol{\Sigma}}_{(p-s) \times (p-s)}$, each exhibiting a first-order autoregressive structure with correlation parameter 0.3; and (3) type III: identity matrix, i.e., $\boldsymbol{\boldsymbol{\Sigma}} = \mathbf{I}_p$. 
\end{case}

\begin{case}\label{case2}
    We set $\boldsymbol{\mu} = (-0.1, 0.2, 0.25, 0.1, -0.2, 0, \ldots, 0)^\top \in \mathbb{R}^p$, $\boldsymbol{\nu} = -\boldsymbol{\mu}$. We also examine three different choices of the covariance \(\boldsymbol{\Sigma}\) as that in the Case \ref{case1}.
\end{case}

The simulations are carried out in both offline and online settings. In the offline setting, we vary the sample sizes \(N\) and feature dimensions \(p\): \((N, p) = (1000, 150)\), \((2000, 400)\), and \((2000, 600)\).
In the online setting, we fix the number of batches \(B = 20\), varying the per-batch sample sizes \(n_b\) and feature dimensions \(p\): $(n_b, p) = (50,150)$, $(100,400)$, $(100,600)$. These choices yield overall sample sizes in the online setting matching those in the offline setting \((1000, 2000, 2000)\), respectively. Thus, it enables a direct comparison between the offline and online smoothed SVM approaches.

We compare our proposed framework with the methods described in \cite{zhang2025renewable}, resulting in the following four estimators: (i) ``offline smooth'': the proposed offline debiased smoothed estimator defined in (\ref{offline debias estimator}); (ii) ``online smooth'': the proposed online debiased smoothed estimator defined in (\ref{online debias estimator}); (iii) ``offline nonsmooth'': the benchmark offline debiased nonsmoothed SVM computed on the full dataset, following the offline implementation in \cite{zhang2025renewable}; (iv) ``online nonsmooth'': the renewable debiased nonsmoothed SVM for streaming data proposed by \cite{zhang2025renewable}.

The simulations are repeated 200 times and assessed by \(10\) different metrics: 

(i) ``abias\_non'': the mean absolute bias of the debiased estimates for the nonzero coefficients, defined as \((1/5)\sum_{j=1}^{5}|\widehat{\boldsymbol{\beta}}^{\mathrm{off}, \mathrm{de}}_j - \boldsymbol{\beta}^*_j|\) in the offline case and \((1/5)\sum_{j=1}^{5}|\widehat{\boldsymbol{\beta}}_{B, j}^{\mathrm{on}, \mathrm{de}} - \boldsymbol{\beta}^*_j|\) in the online case; 

(ii) ``abias\_zero'': the mean absolute bias of the debiased estimates for the zero coefficients;

(iii) ``abias\_all'': the mean absolute bias of the debiased estimates for all coefficients; 

(iv) ``cov\_non'': an average of the empirical coverage probabilities for nonzero coefficients, defined as \((1/5)\sum_{j=1}^{5} 
I\{ \boldsymbol{\beta}^*_j \in \big(\widehat{\boldsymbol{\beta}}^{\mathrm{off},\mathrm{de}}_j - 1.96 \cdot\widehat{{\sigma}}_{j} / \sqrt{n}, \widehat{\boldsymbol{\beta}}^{\mathrm{off},\mathrm{de}}_j  + 1.96 \cdot\widehat{{\sigma}}_{j} / \sqrt{n}\big) \} \) in the offline case and \((1/5)\sum_{j=1}^{5} 
I\{ \boldsymbol{\beta}^*_j \in \big(\widehat{\boldsymbol{\beta}}_{B,j}^{\mathrm{on},\mathrm{de}}  - 1.96 \cdot\widehat{\sigma}_{Bj} / \sqrt{N_B}, \widehat{\boldsymbol{\beta}}_{B,j}^{\mathrm{on},\mathrm{de}}  + 1.96 \cdot\widehat{\sigma}_{Bj} / \sqrt{N_B}\big) \} \) in the online case; 

(v) ``cov\_zero'': an average of the empirical coverage probabilities for zero coefficients; 

(vi) ``cov\_all'': an average of the empirical coverage probabilities for all coefficients; 

(vii) ``len\_non'': an average confidence interval length for nonzero coefficients, defined as 
\( (1/5)\sum_{j=1}^{5} ( 2 \times 1.96 \cdot\widehat{{\sigma}}_{j} / \sqrt{n}) \) in the offline case and \( (1/5)\sum_{j=1}^{5} ( 2 \times 1.96 \cdot\widehat{{\sigma}}_{Bj} / \sqrt{n}) \) in the online case; 

(viii) ``len\_zero'': an average confidence interval length for zero coefficients;

(ix) ``len\_all'': an average confidence interval length for all coefficients; 

(x) ``time'': an average elapsed computing time (in seconds) per replication.

\begin{figure}[htbp]
    \centering
    
    \begin{subfigure}[b]{\textwidth}
        \centering
        \includegraphics[width=0.95\textwidth]{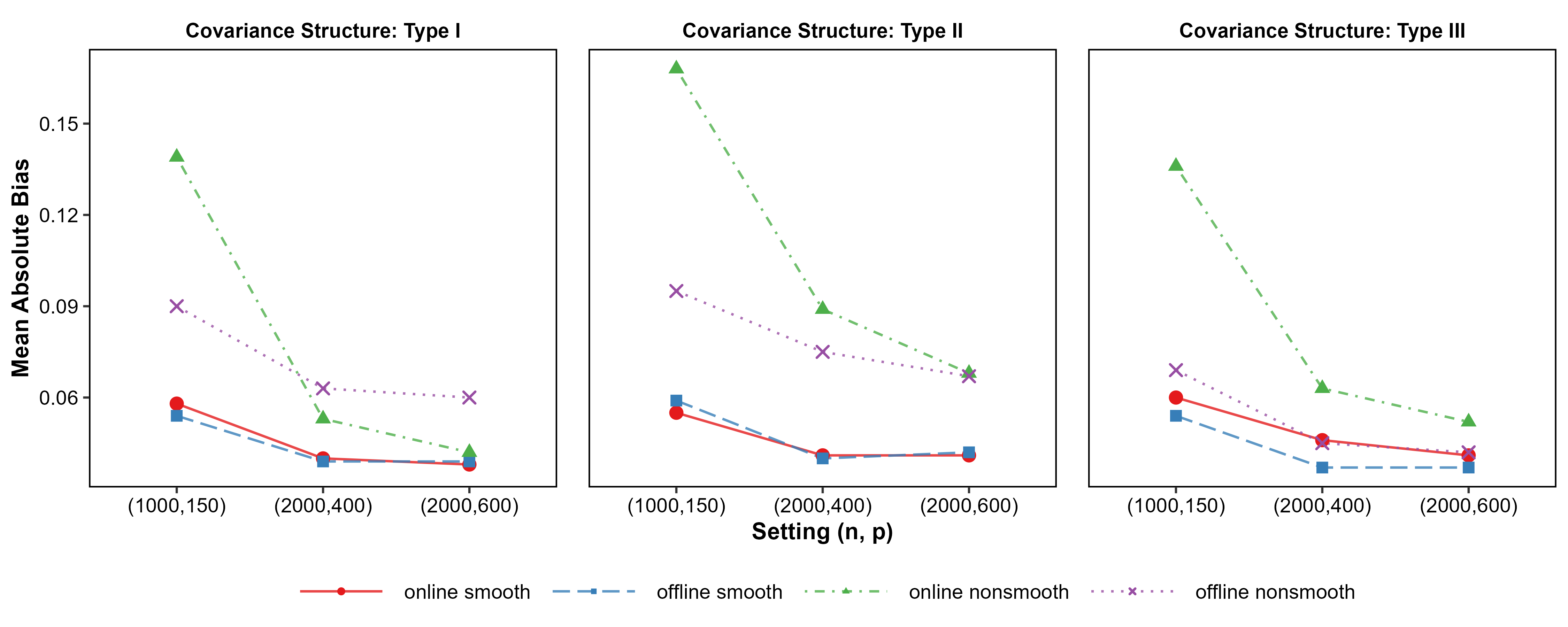}
        \caption{Mean absolute bias for nonzero coefficients}
        \label{fig:case_1_bias}
    \end{subfigure}
    
    \vspace{0.5cm} 

        \begin{subfigure}[b]{\textwidth}
        \centering
        \includegraphics[width=0.95\textwidth]{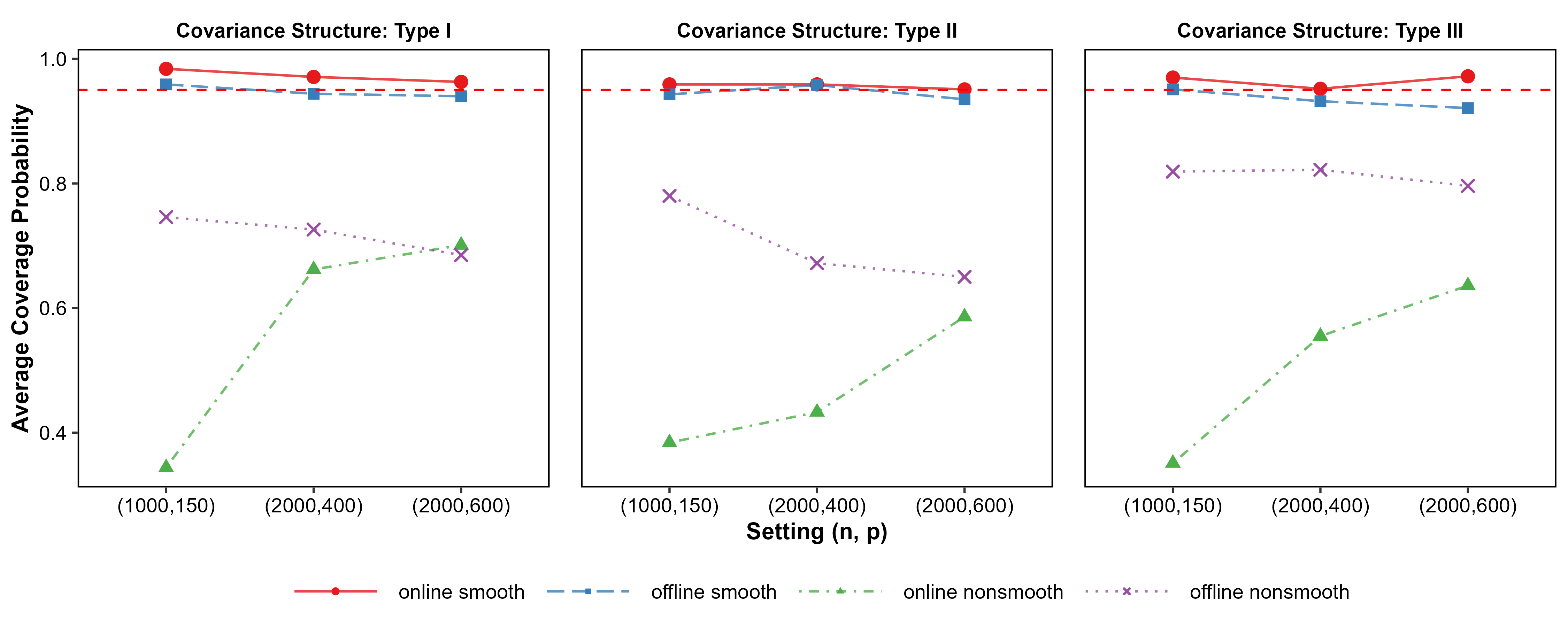}
        \caption{Average coverage probability for nonzero coefficients}
        \label{fig:case_1_coverage}
    \end{subfigure}

    \vspace{0.5cm} 
    
    \begin{subfigure}[b]{\textwidth}
        \centering
        \includegraphics[width=0.95\textwidth]{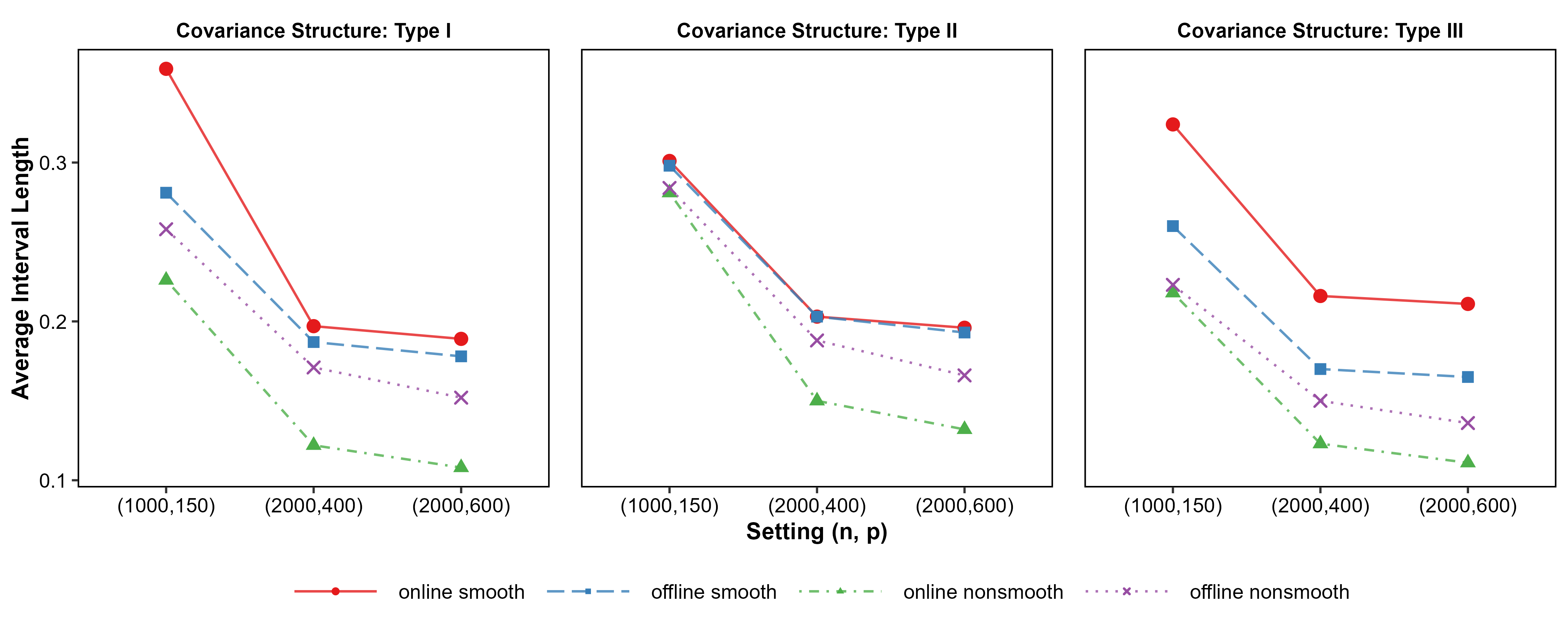}
        \caption{Average interval length for nonzero coefficients}
        \label{fig:case_1_length}
    \end{subfigure}
  
    \caption{Simulation results under Case~\ref{case1} over 200 replications. In the online case, $n = N_B$.}
    \label{fig :case_1_simulation_results}
\end{figure}

\begin{figure}[htbp]
    \centering
    
    \begin{subfigure}[b]{\textwidth}
        \centering
        \includegraphics[width=0.95\textwidth]{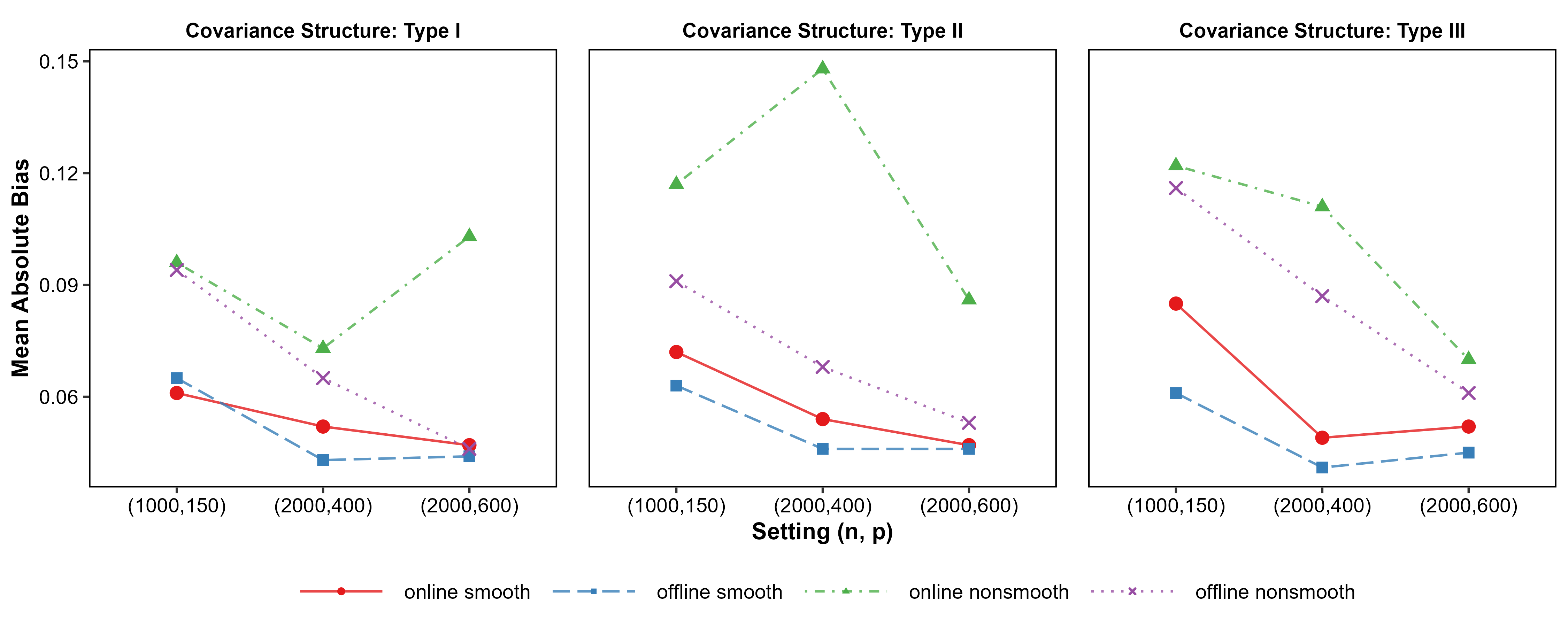}
        \caption{Mean absolute bias for nonzero coefficients}
        \label{fig:case_2_bias}
    \end{subfigure}
    
    \vspace{0.5cm} 

        \begin{subfigure}[b]{\textwidth}
        \centering
        \includegraphics[width=0.95\textwidth]{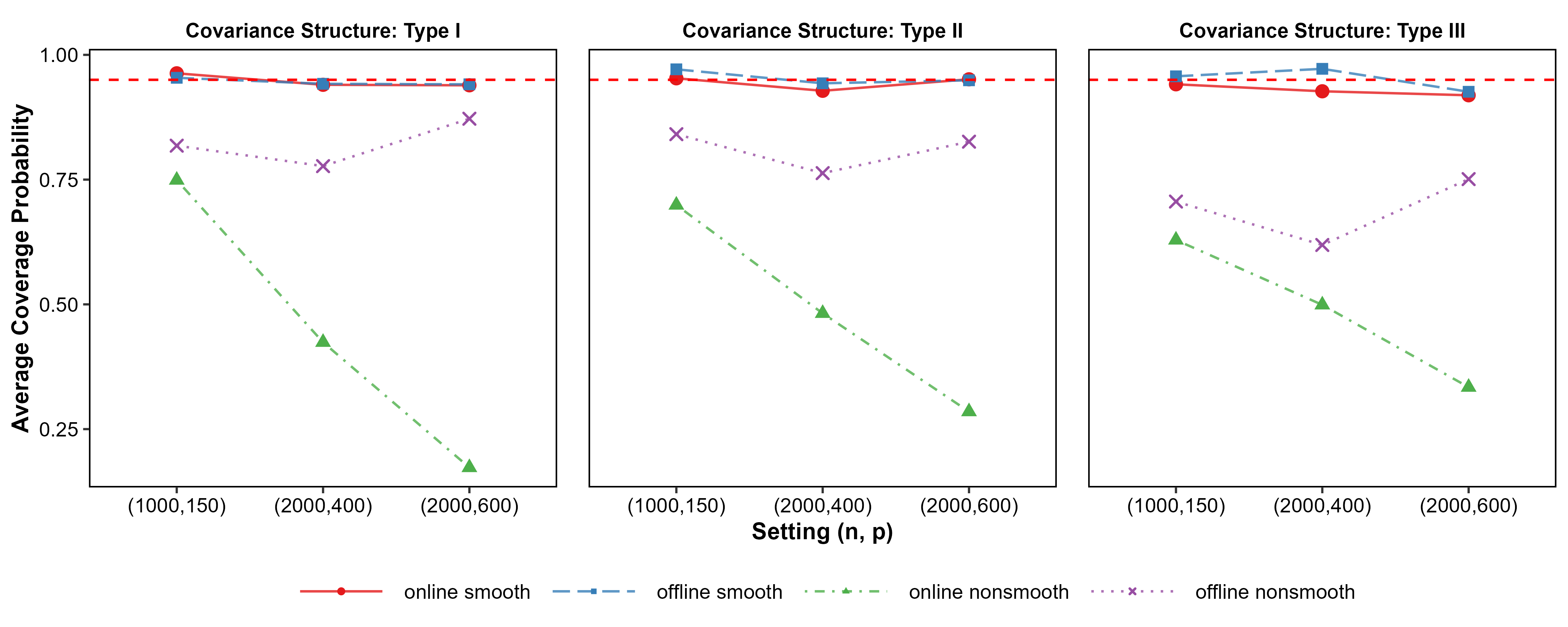}
        \caption{Average coverage probability for nonzero coefficients}
        \label{fig:case_2_coverage}
    \end{subfigure}

    \vspace{0.5cm} 
    
    \begin{subfigure}[b]{\textwidth}
        \centering
        \includegraphics[width=0.95\textwidth]{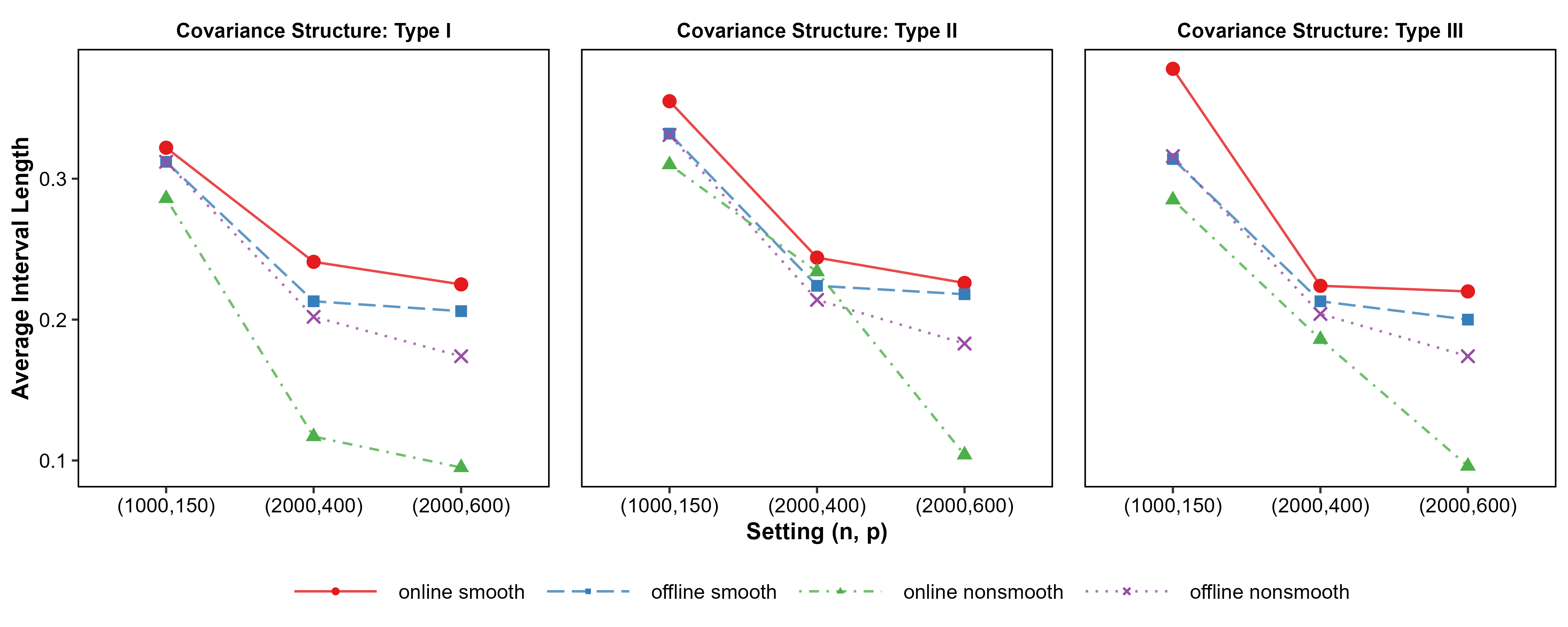}
        \caption{Average interval length for nonzero coefficients}
        \label{fig:case_2_length}
    \end{subfigure}
  
    \caption{Simulation results under Case~\ref{case2} over 200 replications. In the online case, $n = N_B$.}
    \label{fig :case_2_simulation_results}
\end{figure}

We report our simulation results of ``abias\_non'', ``cov\_non'' and ``len\_non'' in Figures~\ref{fig :case_1_simulation_results} and ~\ref{fig :case_2_simulation_results} for Cases \ref{case1} and \ref{case2}, respectively. The remaining metrics are shown in the Appendix \ref{simulation appendix}.

It is evidenced in Figures~\ref{fig :case_1_simulation_results} and \ref{fig :case_2_simulation_results} that the proposed smoothed estimators (red and blue lines) significantly outperform the nonsmoothed ones (green and purple lines). Specifically, the smoothed SVM shows a smaller mean absolute bias for the nonzero coefficients compared to its nonsmoothed counterparts. This demonstrates the ability of the smoothed SVM method to correct shrinkage bias and recover the true magnitude for the nonzero coefficients. Moreover, the empirical coverage probabilities of the smoothed SVM are closely aligned with the nominal 95\% level, whereas the nonsmoothed approaches suffer from under-coverage. Similar performances of convolution smoothing in reducing estimation errors and improving inference validity have also been observed in high-dimensional quantile regression \cite[]{xie2025statistical, he_smooth_quantile}. Furthermore, although our proposed online method utilizes the summary statistics from previous batches rather than the entire dataset, our online smoothed method achieves comparable performance on all metrics to our offline smoothed method. This confirms that the online updating algorithm preserves the estimation accuracy and inference validation without storing historical data. In terms of computation time shown in Figures~\ref{fig:case_1_2_simulation_time} of Appendix \ref{simulation appendix}, the smoothed methods are much faster than the nonsmoothed ones.

\section{Real data analysis \label{application}}
\subsection{Offline data application \label{offline application}}
In this section, we apply our offline debiased method in Section \ref{offline method} to the real-world Large Movie Review Dataset introduced by \cite{imdb_dataset}, which is publicly available at \url{https://ai.stanford.edu/~amaas/data/sentiment/}. The dataset consists of 50,000 highly polarized movie reviews, evenly divided into positive (score $\ge 7/10$) and negative (score $\le 4/10$) classes. To construct the high-dimensional offline dataset, we randomly sample 100 instances each from the positive ($Y=1$) and negative ($Y=-1$) reviews, yielding a total sample size of $N=200$. We select the top $p=500$ most frequent words as features \(\boldsymbol{X}\) and apply the offline debiased estimator in (\ref{offline debias estimator}) and its interval in (\ref{offline interval}) to obtain the coefficients with their corresponding confidence intervals. In Table~\ref{table: offline_features}, we report the results for some representative features, which include both significant ($p < 0.05$) and non-significant words.

\begin{table}[t]
\centering
\caption{Offline debiased estimates and 95\% confidence intervals for representative features}
\label{table: offline_features}
\begin{tabular}{@{}cccc@{}}
\hline
& Feature & Estimate & 95\% Confidence Interval \\
\hline
Significant      &     great      & 0.184  & [0.139, 0.230]  \\
      &     incredible & 0.124  & [0.058, 0.191]  \\
      &     surprise   & 0.103  & [0.050, 0.156]  \\
      &     boring     & $-$0.202 & [$-$0.301, $-$0.103] \\
      &     poor       & $-$0.219 & [$-$0.271, $-$0.168] \\
      &     bad        & $-$0.451 & [$-$0.517, $-$0.385] \\ 
      \hline
Non-significant      &     film       & 0.043  & [$-$0.022, 0.108] \\
     &      actors & 0.114  & [$-$0.036, 0.263] \\
\hline
\end{tabular}
\end{table}

As shown in Table~\ref{table: offline_features}, among the significant words with positive coefficients, ``great'', ``incredible'', and ``surprise'' generally convey favorable sentiments toward the movie, driving the generation of positive reviews. Conversely, words like ``boring'', ``poor'', and ``bad'' typically express disappointment, contributing to negative reviews and displaying negative coefficients. Furthermore, our method successfully identifies non-significant words such as ``film'' and ``actors'', whose confidence intervals include zero, thereby demonstrating the feasibility of our offline method.

\subsection{Streaming data application \label{online application}}
In this section, we apply our online debiased method in Section \ref{online method} to the 10-K Corpus dataset \cite[]{kogan2009predicting}, which is publicly available at \url{https://www.cs.cmu.edu/~ark/10K/}. The corpus pairs the annual 10-K reports of thousands of U.S. public companies (1996--2006) with their corresponding pre- and post-publication stock return volatilities over 12 months. We analyze a total of $N_B = 16,794$ annual financial reports published between 2002 and 2006, and partition the observations into batches at three-month intervals with a total batch size of $B = 20$. The response variable $Y$ is defined as a binary indicator, where $Y = 1$ if the stock volatility in the twelve months after the report increases compared to the twelve months before, and $Y = -1$ otherwise. In other words, \(Y=1\) represents an increase in risk regarding the company’s future, while $Y = -1$ represents a decrease in risk. The high-dimensional feature vector $\boldsymbol{X}$ consists of the top $p = 500$ most frequent phrases (including both unigrams and bigrams) extracted from the report texts.

\begin{figure}
    \centering
    \includegraphics[width=1\linewidth]{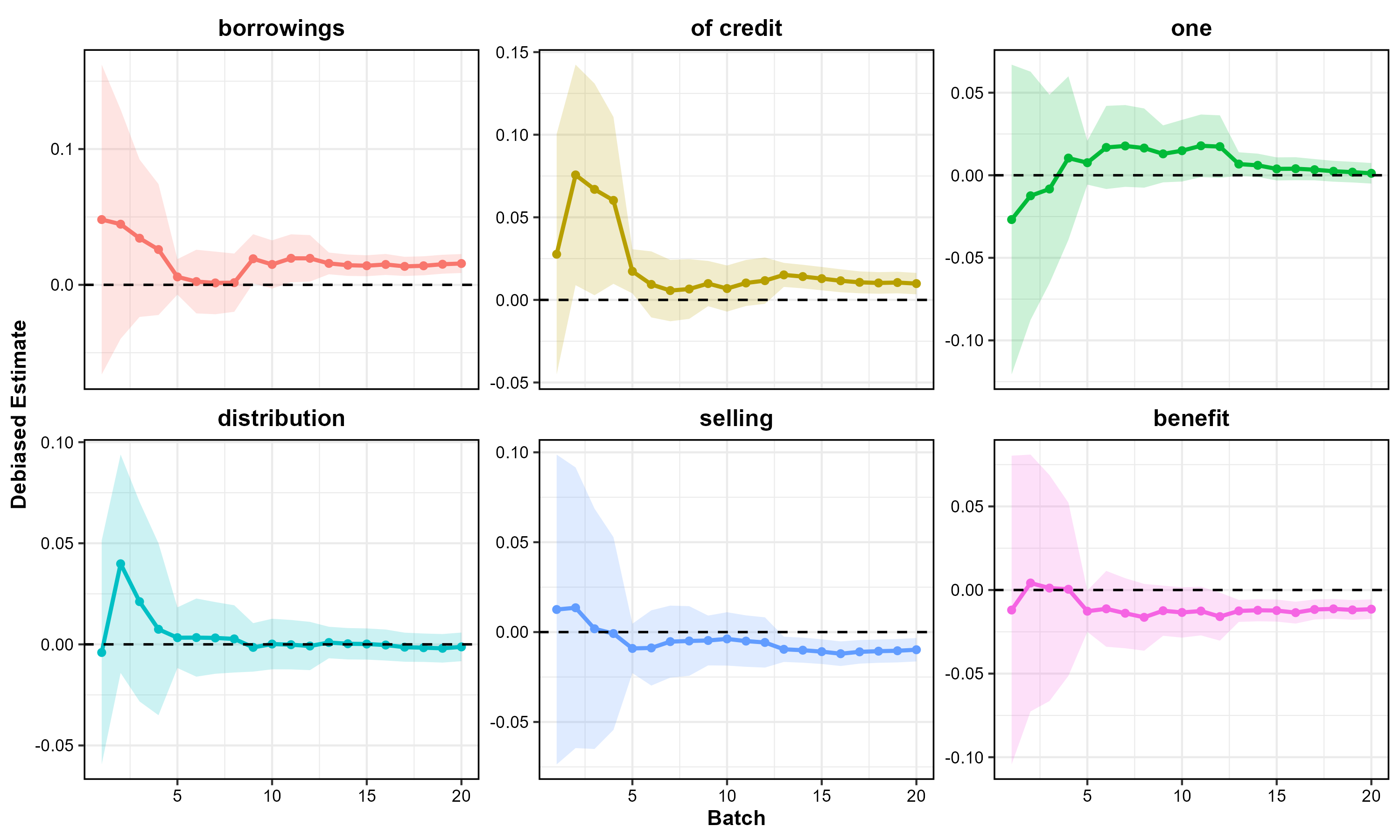}
    \caption{Trajectories of the online debiased estimates (solid lines) and corresponding 95\% confidence intervals (shaded areas) for six features.}
    \label{fig:online trajectory}
\end{figure}

Using the online debiased estimator in (\ref{online debias estimator}) and the interval construction in (\ref{online interval}), we track the estimation trajectories and 95\% confidence intervals of six features: ``borrowings'', ``of credit'', ``one'', ``distribution'', ``selling'', and ``benefit'' in Figure~\ref{fig:online trajectory}. We observe that as more data arrives, the point estimates become stable, and their confidence intervals become much narrower. This shows that our online method can continuously improve accuracy and reduce uncertainty.

As illustrated in Figure~\ref{fig:online trajectory}, the estimated coefficients for these features align well with financial intuition. Specifically, the words ``borrowings'' and ``of credit'' have positive coefficients. These words are directly related to corporate debt. When a company uses them frequently, it usually points to a worse financial situation. This makes investors nervous and increases stock volatility. In contrast, ``selling'' and ``benefit'' have negative coefficients. These words suggest higher income and profits, which state the stock is less likely to fluctuate. Finally,  our method can effectively identify words that have no real impact on stock volatility. Specifically, our method correctly shows that ``one'' and ``distribution'' are not significant. Their estimates stay close to zero, and their confidence intervals consistently cover zero. 

\section{Discussion \label{discussion}}
In this paper, we present a debiased Lasso methodology for constructing confidence intervals of the smoothed SVM in an offline setting. In addition, we develop an online updating debiased Lasso procedure tailored to streaming data, in which the updating scheme relies solely on summary statistics from previous batches. We further establish that the confidence intervals of our offline and online methods are asymptotically valid. Simulation studies and real data applications demonstrate that our methods perform effectively for statistical inference in both offline and online settings.

In this paper, we focus on the statistical inference for any single parameter of the decision boundary. In some applications, multiple parameters are of interest. For example, one may wish to jointly infer the parameters corresponding to several features to assess their combined effect on the orientation and position of the decision boundary, or to test hypotheses about groups of parameters rather than individual components \cite[]{multiple_infer_1, multiple_infer_2}. In such cases, the joint asymptotic distribution of the debiased estimate is required. Our method is readily extendable to this case. As an illustrative example, we consider the offline debiased estimator \(\widehat{\boldsymbol{\beta}}^{\mathrm{off,de}}\). Theorem \ref{theorem: offline asymptotic} indicates that \(\sqrt{n}(\widehat{\boldsymbol{\beta}}^{\mathrm{off},\mathrm{de}} - \boldsymbol{\beta}^*) = Z^{\mathrm{off}} + W^{\mathrm{off}}\), where \(W^{\mathrm{off}}=(W^{\mathrm{off}}_1,...,W^{\mathrm{off}}_{p+1})^T\) is asymptotically ignorable, and \(Z^{\mathrm{off}}=(Z^{\mathrm{off}}_1,...,Z^{\mathrm{off}}_{p+1})^T\) is asymptotically normal with mean \(E(Z^{\mathrm{off}})=0\) and the covariance \(\mathrm{Cov}(Z^{\mathrm{off}})= (\boldsymbol{\Theta}^*)^\top \boldsymbol{\Sigma}^* \boldsymbol{\Theta}^*\), with a similar discussion under Theorem \ref{theorem: offline asymptotic}. By utilizing the asymptotic normality of \(\sqrt{n}(\widehat{\boldsymbol{\beta}}^{\mathrm{off},\mathrm{de}} - \boldsymbol{\beta}^*)\), we are able to perform joint inference on the parameters defining the decision boundary.

In this paper, the response \(Y\) is assumed binary, e.g., \(Y\in\{-1,1\}\). In certain applications, the response \(Y\) may be multi-category \cite[]{multiple_classification}. We may utilize the one-versus-rest or one-versus-one strategies to perform our method and therefore construct confidence intervals for the parameters of the decision boundaries. The one-versus-rest approach fits a separate binary classifier for each category against all remaining categories, while the one-versus-one approach constructs classifiers for all pairs of categories. In both cases, our inference procedure can be applied to each binary subproblem.

\begin{appendix}
    \section{Additional simulation results \label{simulation appendix}}
\begin{figure}[htbp]
    \centering

        \begin{subfigure}[b]{\textwidth}
        \centering
        \includegraphics[width=0.95\textwidth]{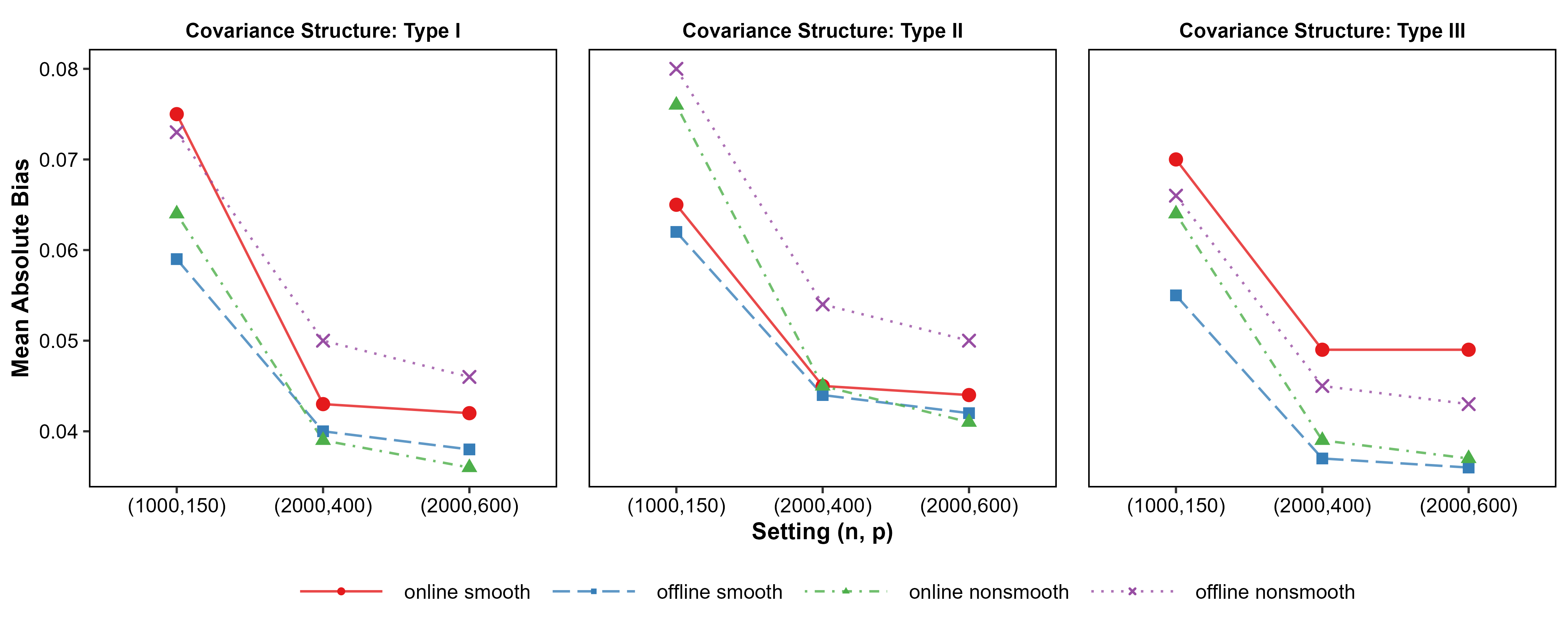}
        \caption{Mean absolute bias for zero coefficients} 
        \label{fig:case_1_zero_abias}
    \end{subfigure}

    \vspace{0.5cm}

        \begin{subfigure}[b]{\textwidth}
        \centering
        \includegraphics[width=0.95\textwidth]{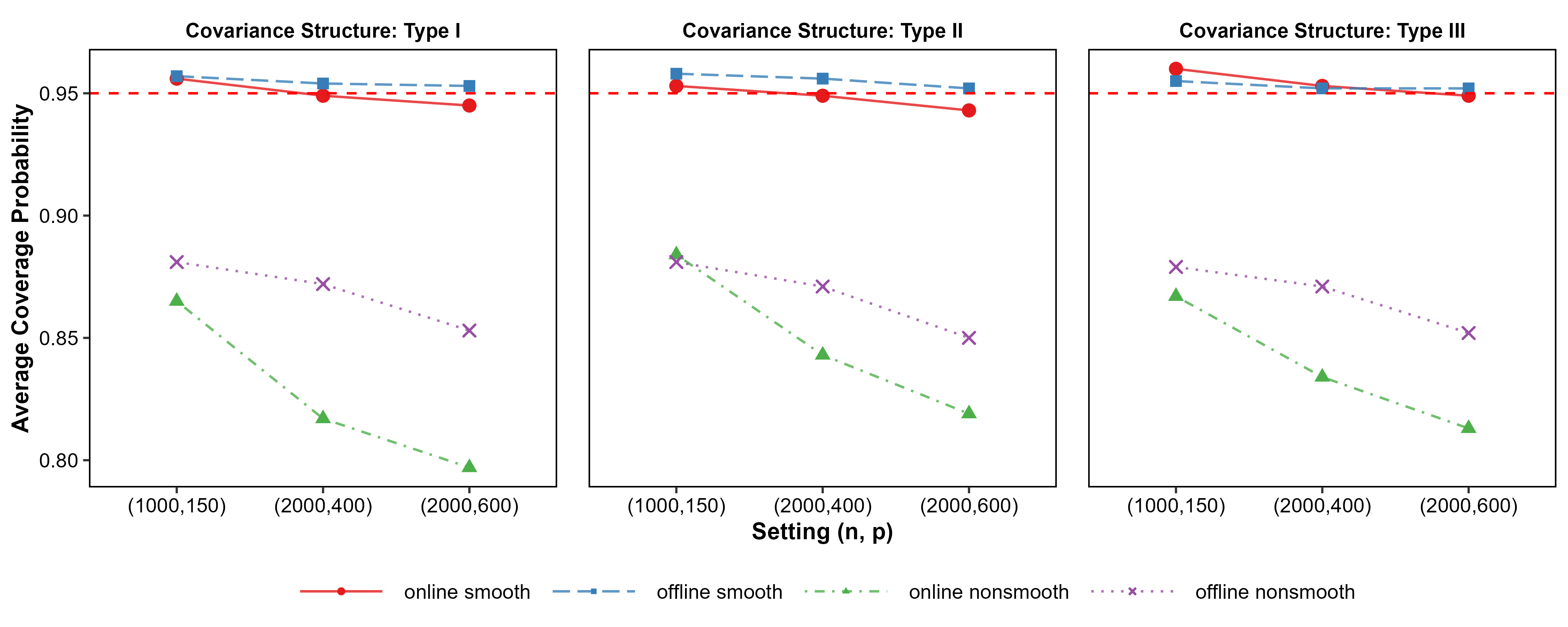}
        \caption{Average coverage probability for zero coefficients} 
        \label{fig:case_1_zero_cov}
    \end{subfigure}

    \vspace{0.5cm}

        \begin{subfigure}[b]{\textwidth}
        \centering
        \includegraphics[width=0.95\textwidth]{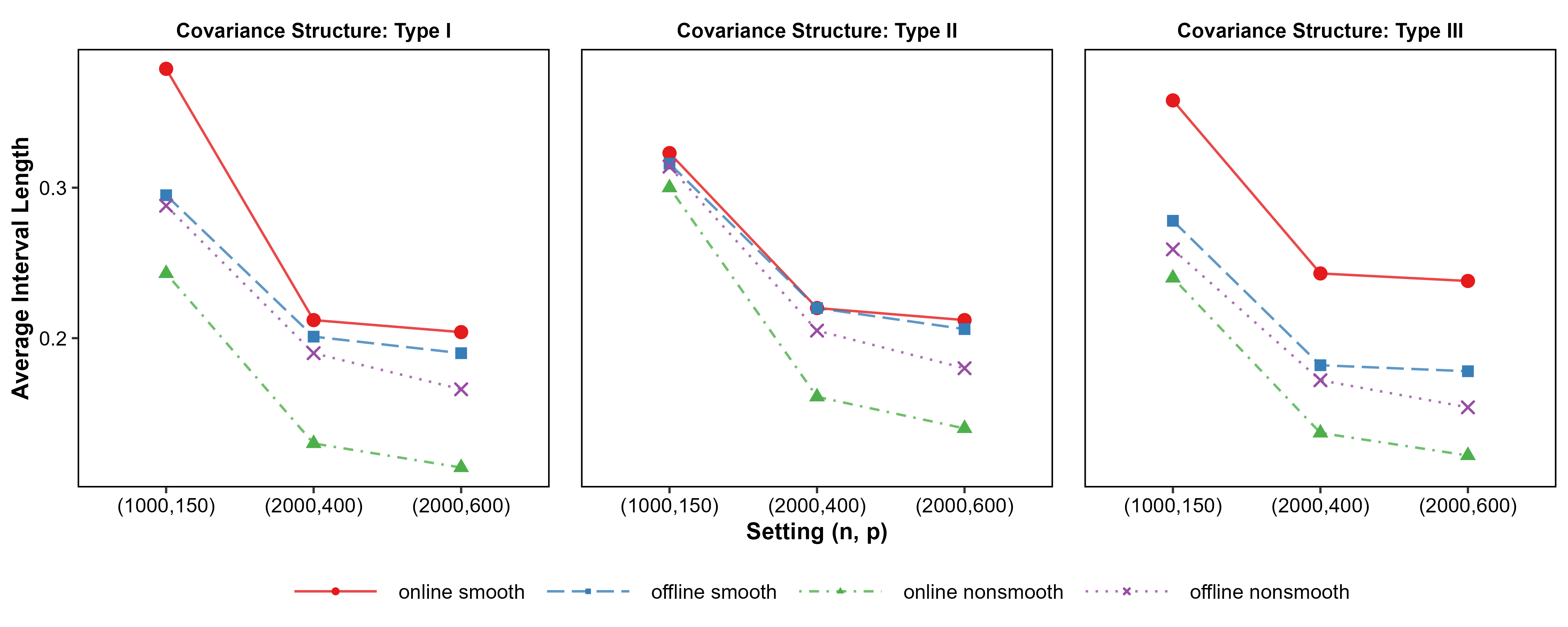}
        \caption{Average interval length for zero coefficients} 
        \label{fig:case_1_zero_length}
    \end{subfigure}
    
    \caption{The results of zero coefficients under Case~\ref{case1} over 200 replications. In the online case, $n = N_B$.}
    \label{fig:case_1_simulation_zero}
\end{figure}

\begin{figure}[htbp]
    \centering

        \begin{subfigure}[b]{\textwidth}
        \centering
        \includegraphics[width=0.95\textwidth]{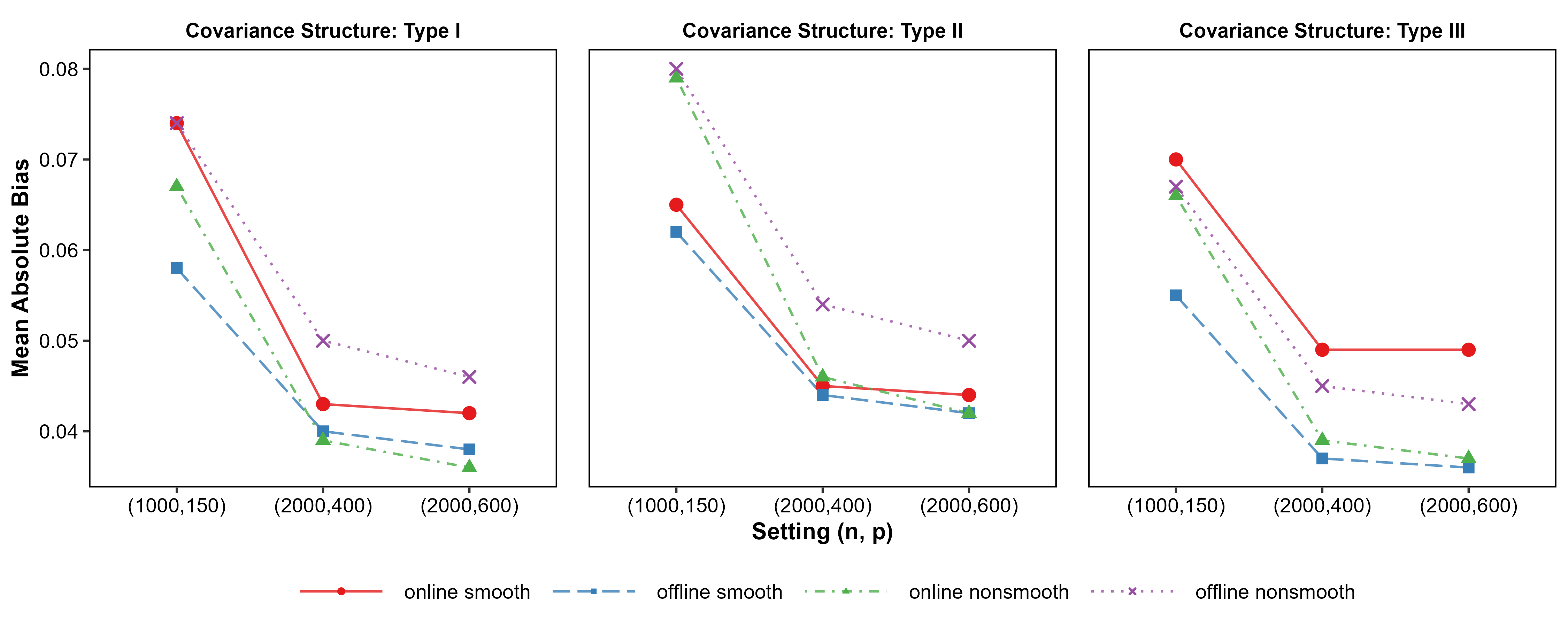}
        \caption{Mean absolute bias for all coefficients} 
        \label{fig:case_1_all_abias}
    \end{subfigure}

    \vspace{0.5cm}

        \begin{subfigure}[b]{\textwidth}
        \centering
        \includegraphics[width=0.95\textwidth]{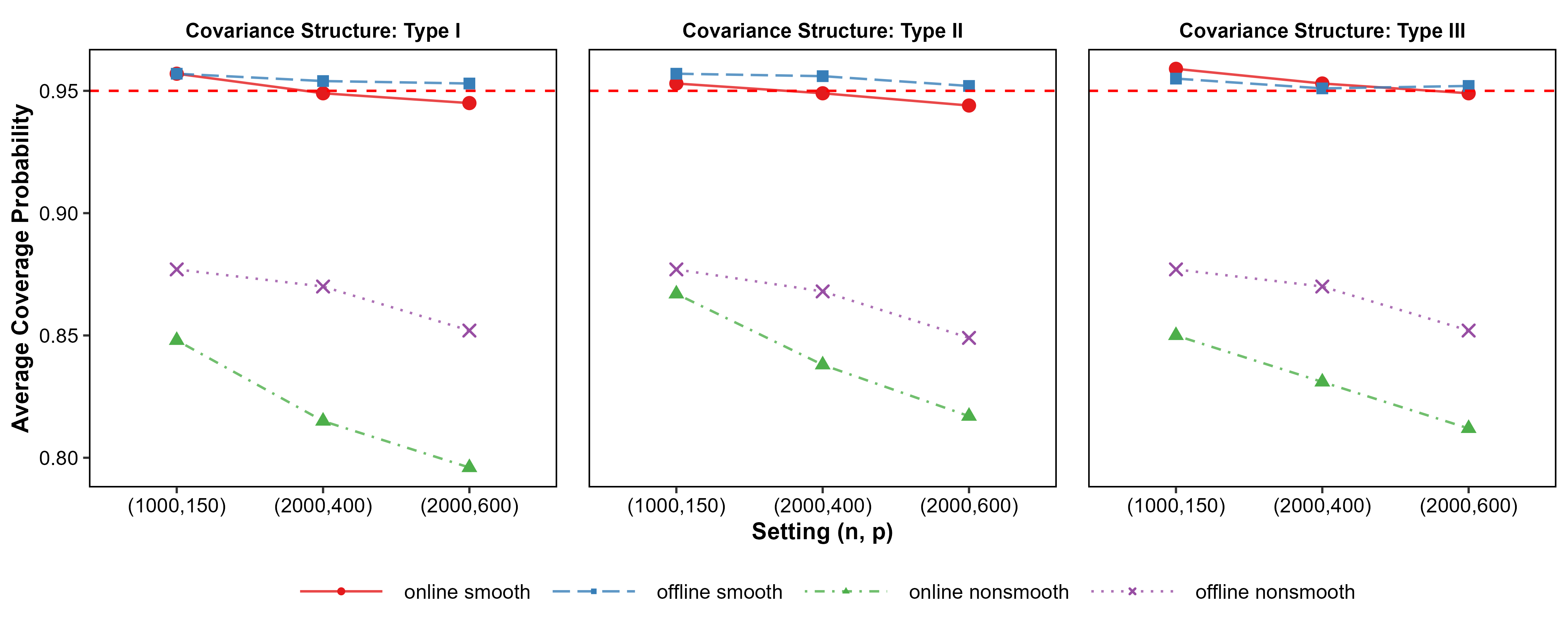}
        \caption{Average coverage probability for all coefficients} 
        \label{fig:case_1_all_cov}
    \end{subfigure}

    \vspace{0.5cm}

        \begin{subfigure}[b]{\textwidth}
        \centering
        \includegraphics[width=0.95\textwidth]{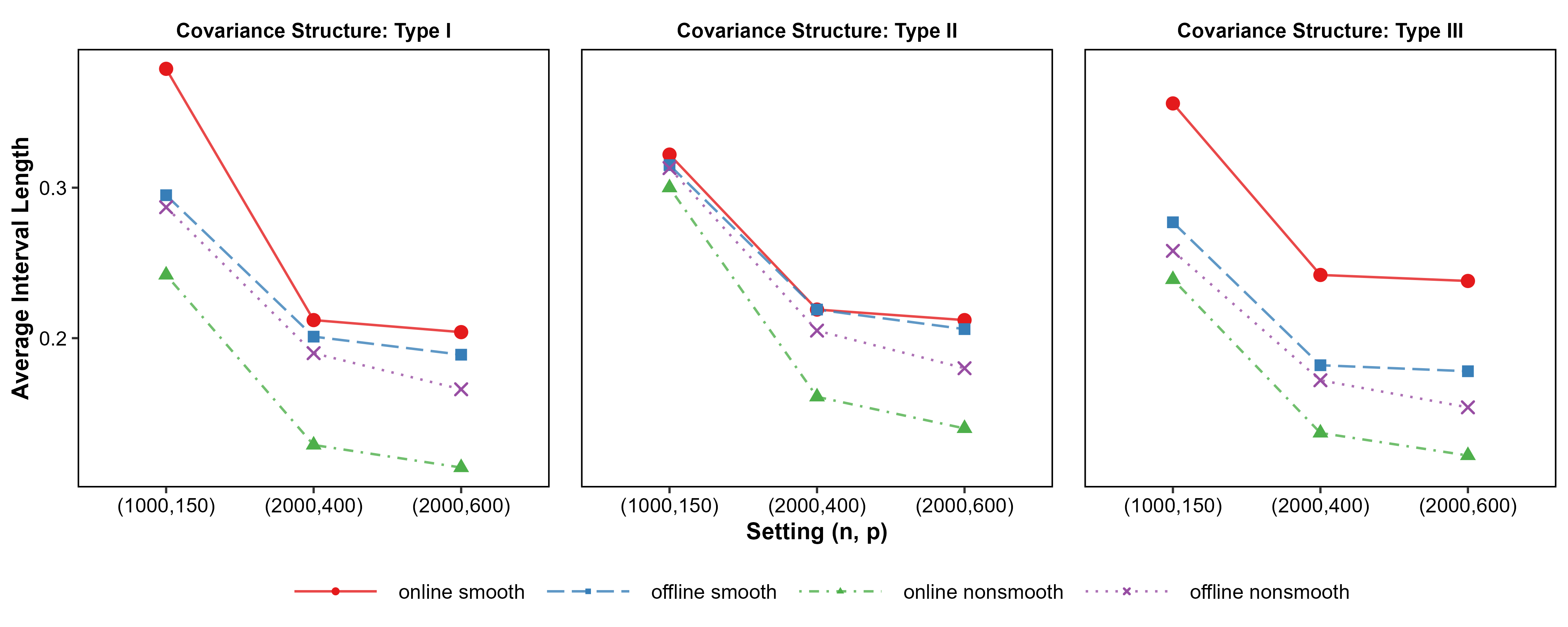}
        \caption{Average interval length for all coefficients} 
        \label{fig:case_1_all_length}
    \end{subfigure}
    
    \caption{The results of all coefficients under Case~\ref{case1} over 200 replications. In the online case, $n = N_B$.}
    \label{fig:case_1_simulation_all}
\end{figure}

\begin{figure}[htbp]
    \centering

        \begin{subfigure}[b]{\textwidth}
        \centering
        \includegraphics[width=0.95\textwidth]{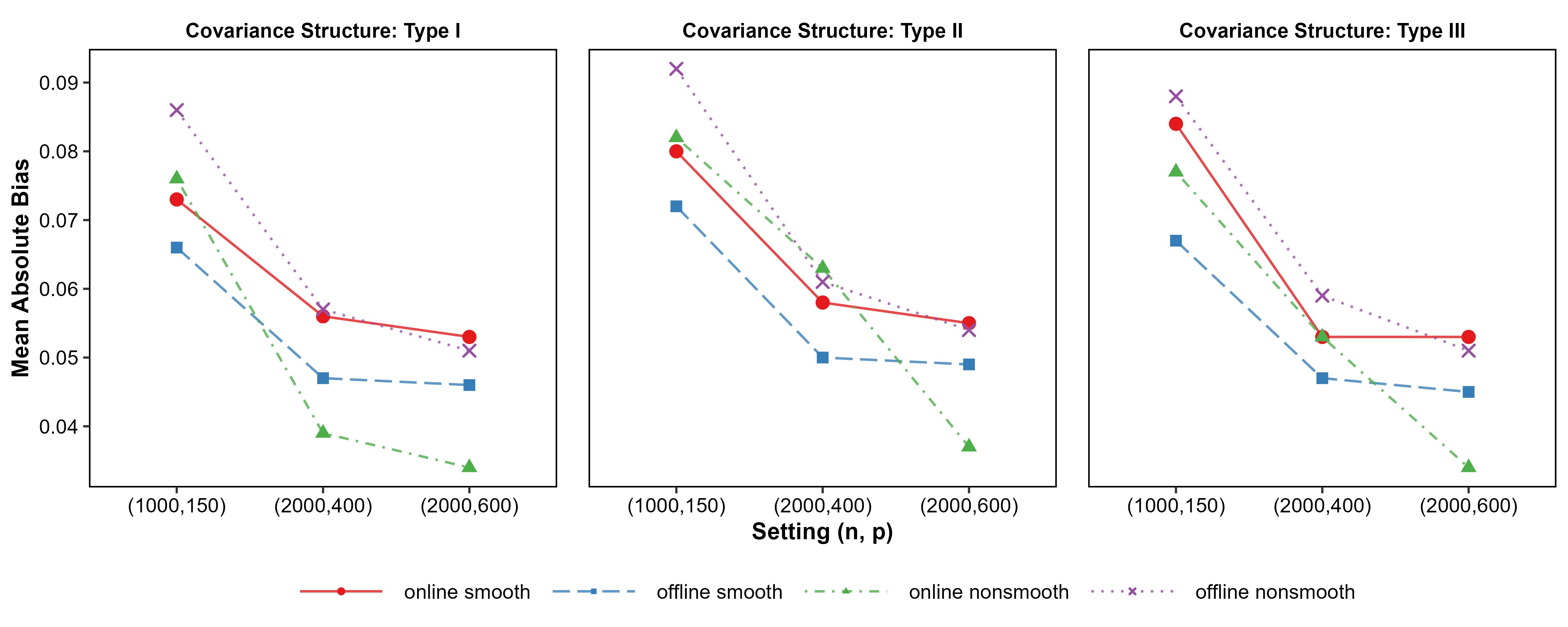}
        \caption{Mean absolute bias for zero coefficients} 
        \label{fig:case_2_zero_abias}
    \end{subfigure}

    \vspace{0.5cm}

        \begin{subfigure}[b]{\textwidth}
        \centering
        \includegraphics[width=0.95\textwidth]{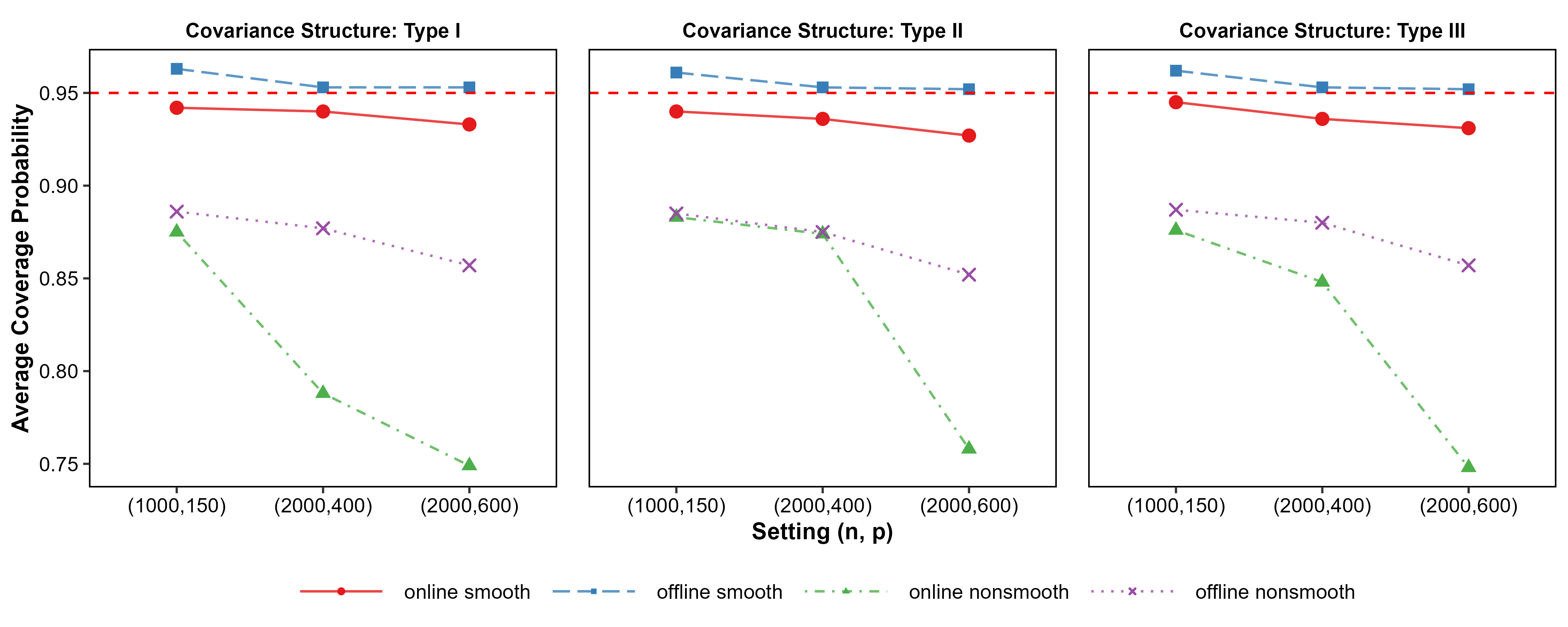}
        \caption{Average coverage probability for zero coefficients} 
        \label{fig:case_2_zero_cov}
    \end{subfigure}

    \vspace{0.5cm}

        \begin{subfigure}[b]{\textwidth}
        \centering
        \includegraphics[width=0.95\textwidth]{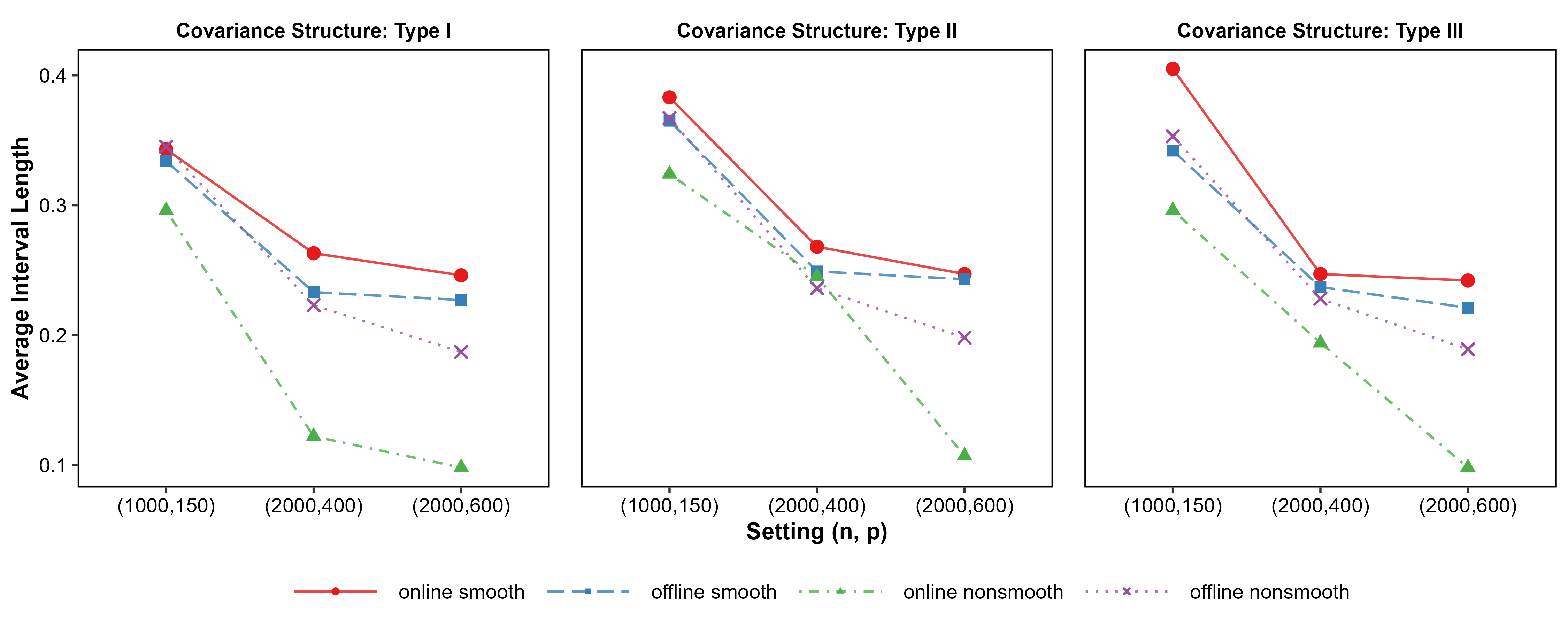}
        \caption{Average interval length for zero coefficients} 
        \label{fig:case_2_zero_length}
    \end{subfigure}
    
    \caption{The results of zero coefficients under Case~\ref{case2} over 200 replications. In the online case, $n = N_B$.}
    \label{fig:case_2_simulation_zero}
\end{figure}

\begin{figure}[htbp]
    \centering

        \begin{subfigure}[b]{\textwidth}
        \centering
        \includegraphics[width=0.95\textwidth]{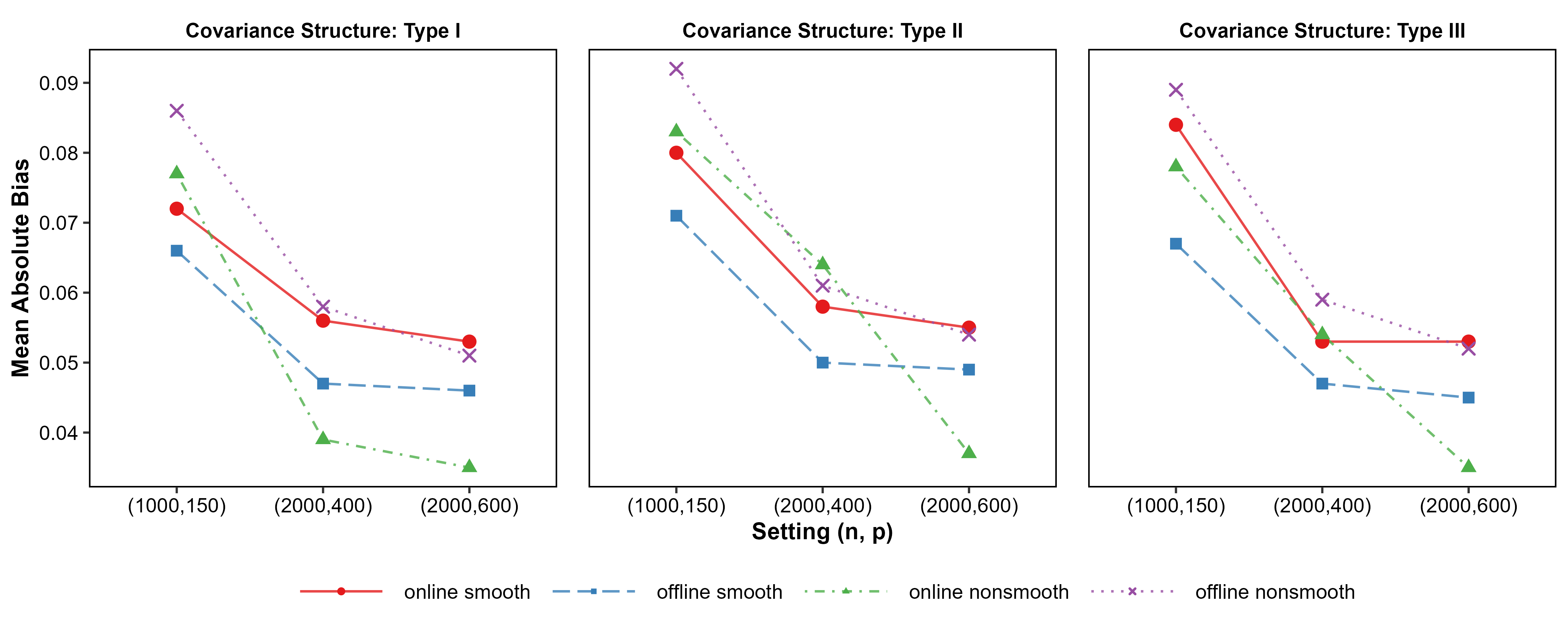}
        \caption{Mean absolute bias for all coefficients} 
        \label{fig:case_2_all_abias}
    \end{subfigure}

    \vspace{0.5cm}

        \begin{subfigure}[b]{\textwidth}
        \centering
        \includegraphics[width=0.95\textwidth]{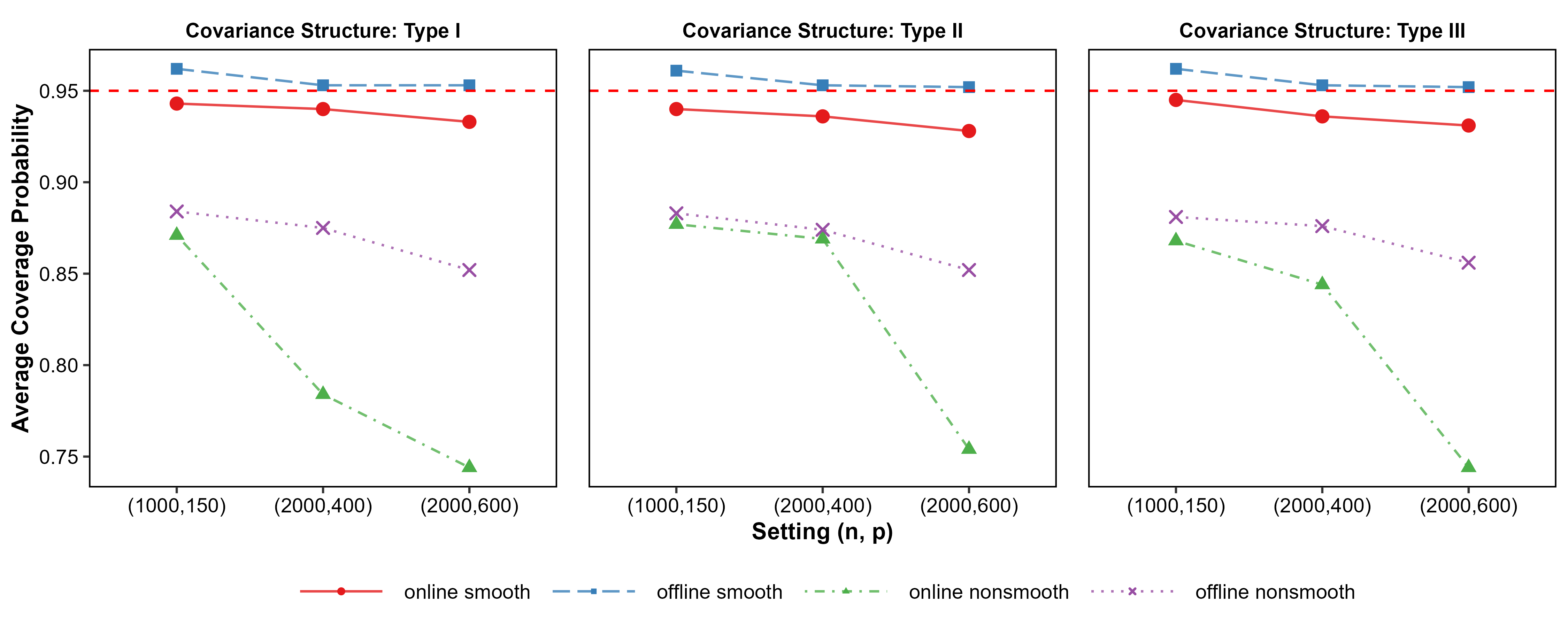}
        \caption{Average coverage probability for all coefficients} 
        \label{fig:case_2_all_cov}
    \end{subfigure}

    \vspace{0.5cm}

        \begin{subfigure}[b]{\textwidth}
        \centering
        \includegraphics[width=0.95\textwidth]{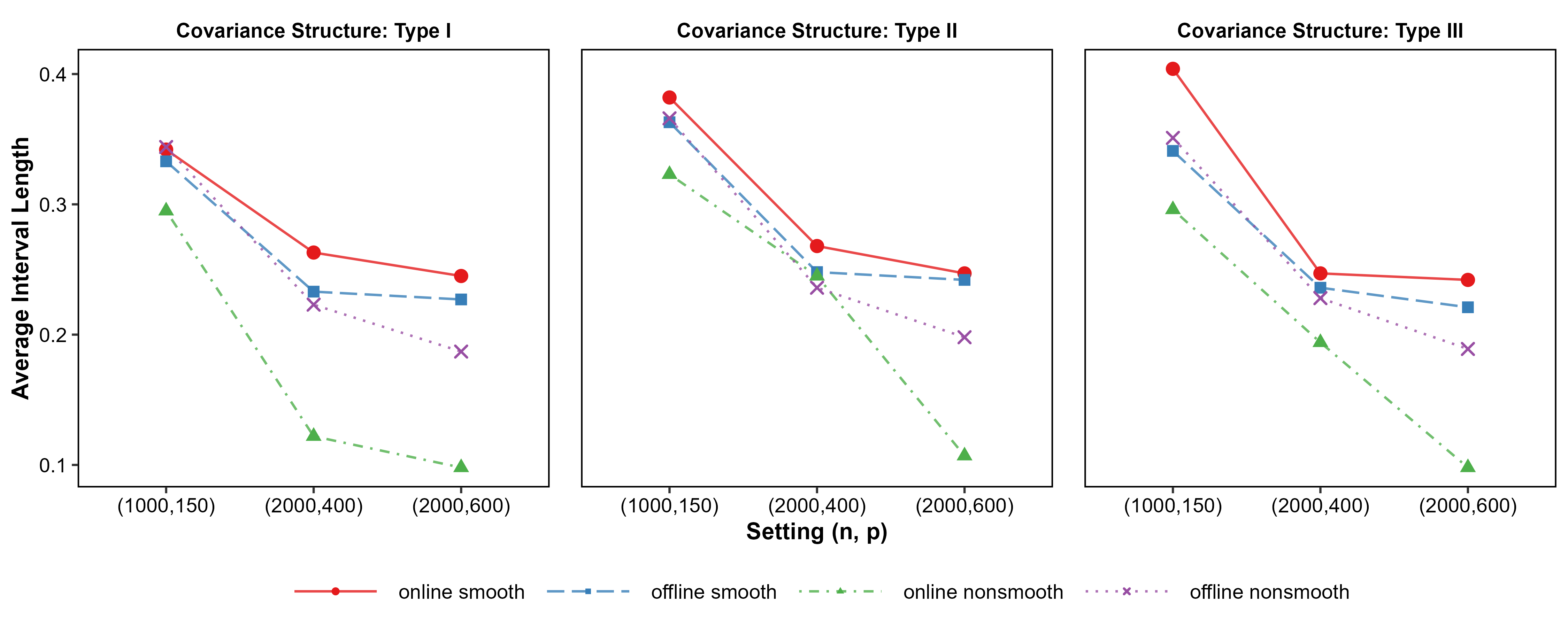}
        \caption{Average interval length for all coefficients} 
        \label{fig:case_2_all_length}
    \end{subfigure}
    
    \caption{The results of all coefficients under Case~\ref{case2} over 200 replications. In the online case, $n = N_B$.}
    \label{fig:case_2_simulation_all}
\end{figure}

\begin{figure}[ht]
    \centering
    
    \begin{subfigure}[b]{\textwidth}
        \centering
        \includegraphics[width=0.95\textwidth]{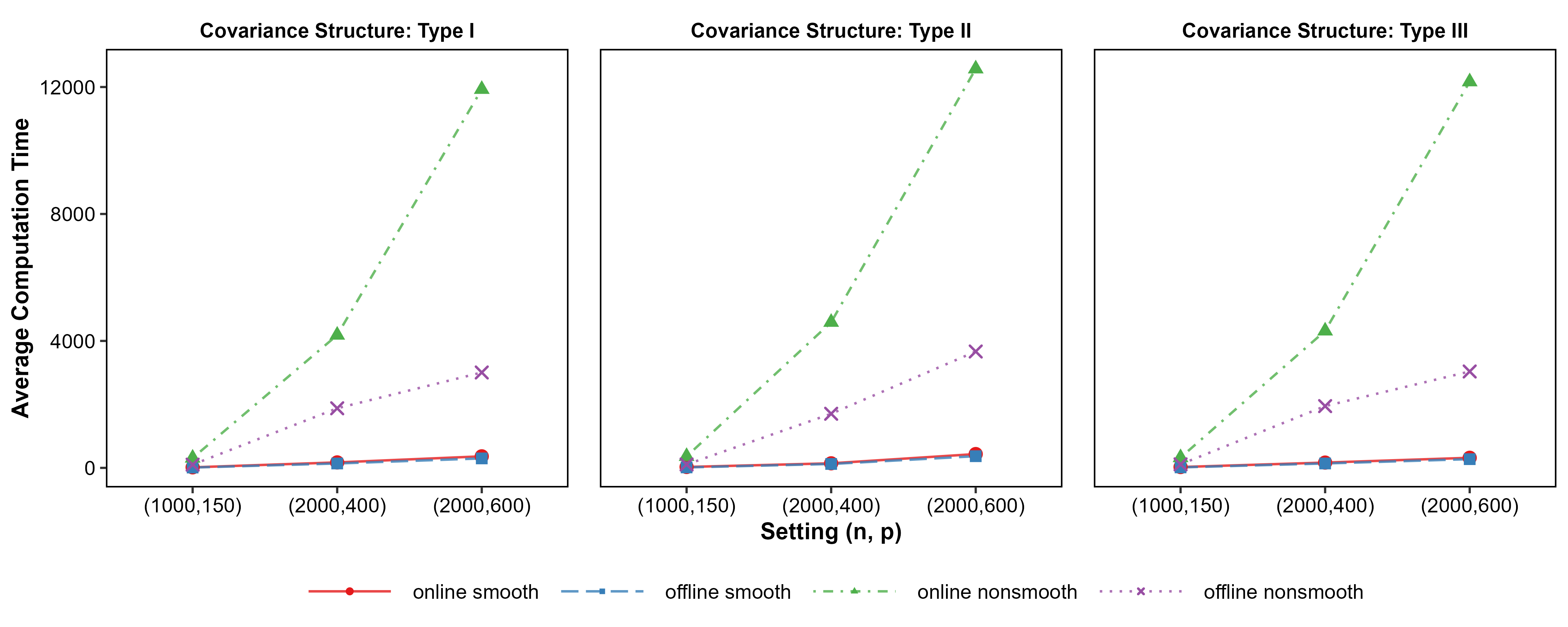}
        \caption{Average computation time under Case~\ref{case1}} 
        \label{fig:case_1_time}
    \end{subfigure}
    
    \vspace{0.5cm} 

        \begin{subfigure}[b]{\textwidth}
        \centering
        \includegraphics[width=0.95\textwidth]{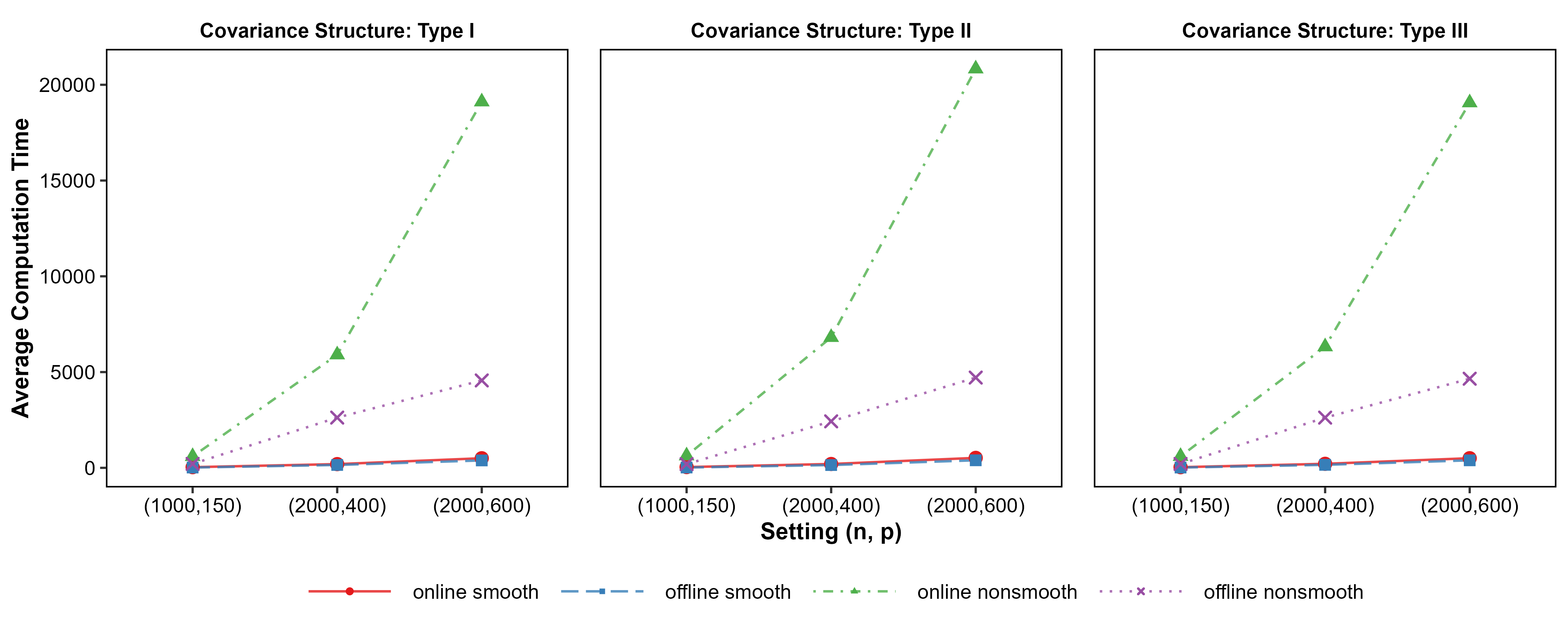}
        \caption{Average computation time under Case~\ref{case2}} 
        \label{fig:case_2_time}
    \end{subfigure}
    
    \caption{Average computation time under Cases~\ref{case1} and \ref{case2} over 200 replications. In the online case, $n = N_B$.}
    \label{fig:case_1_2_simulation_time}
\end{figure}

    In this appendix, we provide additional metrics to evaluate our methods. These metrics include: `abias\_zero'',  ``abias\_all'', ``cov\_zero'', ``cov\_all'', ``len\_zero'', ``len\_all'' and ``time'' as detailed in Section \ref{simulation}. The results are reported in Figures~\ref{fig:case_1_simulation_zero}--\ref{fig:case_1_2_simulation_time}. As illustrated in Figures~\ref{fig:case_1_zero_abias}, \ref{fig:case_1_all_abias}, \ref{fig:case_2_zero_abias}, and \ref{fig:case_2_all_abias}, the mean absolute bias of the smoothed estimator decreases as the dimension $p$ grows, and the offline smoothed method consistently achieves a lower bias than the offline nonsmoothed method. Figures~\ref{fig:case_1_zero_cov}, \ref{fig:case_1_all_cov}, \ref{fig:case_2_zero_cov} and \ref{fig:case_2_all_cov} report the empirical coverage probability. The coverage probability of the smoothed method is close to the 95\% nominal level, whereas the nonsmoothed method exhibits under-coverage. Moreover, as we can see in Figures~\ref{fig:case_1_zero_length}, \ref{fig:case_1_all_length}, \ref{fig:case_2_zero_length} and \ref{fig:case_2_all_length}, the average interval length of the smoothed method shrinks as the sample size $n$ increases. Furthermore, Figure~\ref{fig:case_1_2_simulation_time} shows that the smoothed method is computationally faster than the nonsmoothed one. 

    In summary, these results demonstrate that the smoothing technique yields significantly faster computation and superior inference over the nonsmoothed methods, while the online smoothed method successfully matches the efficiency and accuracy of the offline smoothed one.

\end{appendix}

\bibliography{reference}      

\end{document}